\documentclass[preprint,amsmath,amssymb,prl]{revtex4}
\usepackage{graphicx}
\usepackage{amssymb}

\begin{document}
\bibliographystyle{aip}

\title{Theory of spatial coherence in near-field Raman scattering}

\author{Luiz Gustavo Can\c{c}ado,$^{1}$ Ryan Beams,$^{2}$ Ado Jorio,$^{1}$ and Lukas Novotny$^{3}$}

\affiliation{$^1$Departamento de F\'{\i}sica, Universidade Federal de Minas Gerais, Belo Horizonte, MG 30123-970, Brazil.\\
$^2$ Institute of Optics, University of Rochester, Rochester, NY 14627, USA.\\
$^3$ Photonics Laboratory, ETH Z{\"u}rich, 8093 Z{\"u}rich, Switzerland.}


\begin{abstract}
A theoretical study describing the coherence properties of near-field Raman scattering in two- and one-dimensional systems is presented. The model is applied to the Raman modes of pristine graphene and graphene edges. Our analysis is based on the tip-enhanced Raman scheme, in which a sharp metal tip located near the sample surface acts as a broadband optical antenna that transfers the information contained in the spatially-correlated (but non-propagating) near-field to the far-field. The dependence of the scattered signal on the tip-sample separation is explored, and the theory predicts that the signal enhancement depends on the particular symmetry of a vibrational mode. The model can be applied to extract of the correlation length $L_{\rm c}$  of optical phonons from experimentally recorded near-field Raman measurements. Although the coherence properties of optical phonons have been broadly explored in the time and frequency domains, the spatially-resolved approach presented here provides an alternative probe for the study of local material properties at the nanoscale.
\end{abstract}

\date{Submitted on \today}

\pacs{ }

\maketitle

Raman scattering in crystals is usually treated in the literature as a spatially incoherent process~\cite{cardonaII,loudonbook}. In other words, the scattered field from different sample points is considered to be spatially uncorrelated. This approach is supported by the early theory of coherence stating that the field emitted by an incoherent source at a given wavelength $\lambda$ is spatially uncorrelated on length-scales larger than $\lambda/2$ (measured from the surface of the scatterer)~\cite{wolf1975}. As a consequence, correlations on length-scales smaller than $\lambda/2$ are inaccessible in standard light scattering and the signal recorded in the far-field is incoherent. With the advent of near-field optics and nanoscience  in general, studies of thermal emitters revealed correlation lengths much shorter than $\lambda$~\cite{greffet1999,greffet2000,wolf2003,apostol2003,wolf2004}. Here we show that similar effects underlie near-field Raman scattering and that correlation lengths much smaller than $\lambda/2$ can be extracted from measured data. Thus, near-field Raman scattering must take into account subwavelength correlations and associated interference effects.\\

The coherence of lattice vibrations is of particular importance for graphene-based electronics, since the scattering of optical phonons provides the main channel for relaxation of charge carriers and heat dissipation in this material system~\cite{kang2010,heinz2010,park2010,wu2012,sun2012,saito2013}. In this work, we derive a theory for near-field Raman scattering in one- and two-dimensional systems, and apply the theory to pristine graphene and graphene edges. Our analysis is based on the tip-enhanced Raman scattering (TERS) scheme, in which a sharp metal tip is located near the sample at distances  much smaller than $\lambda$. The tip acts as a broadband optical antenna, transferring the information contained in the spatially-correlated (but non-propagating) near-field to the far-field. We analyze the dependence of the scattered signal on the tip-sample separation distance, and show that different vibrational modes (with distinct symmetries and dimensionalities) give rise to different tip-sample distance dependencies. The theory has been used to measure the correlation length $L_{\rm c}$ of optical phonons in graphene, for which we found $L_{\rm c}\approx30$\,nm~\cite{beams2014}. Although the correlation properties of optical phonons have been broadly explored in the time and frequency domains~\cite{kurz2000}, the spatially-resolved approach presented here provides an alternative probe for the study of local material properties at the nanoscale.\\

Raman scattering is an inelastic scattering process where the incident and scattered photons present different energies. The energy difference is equal to the energy of a quantum of vibration (phonon) that is either created (Stokes Raman component) or annihilated (anti-Stokes Raman component)~\cite{cardonaII,loudonbook}. The scope of the present study is to extract the correlation length of phonons in crystals by exploring the coherence properties of the inelastically scattered field in Raman processes. In order to introduce the theory and the parameters involved, we briefly discuss the classical theory of light scattering, keeping the focus on the spatial domain.\\

\begin{figure}
\includegraphics[scale=0.6]{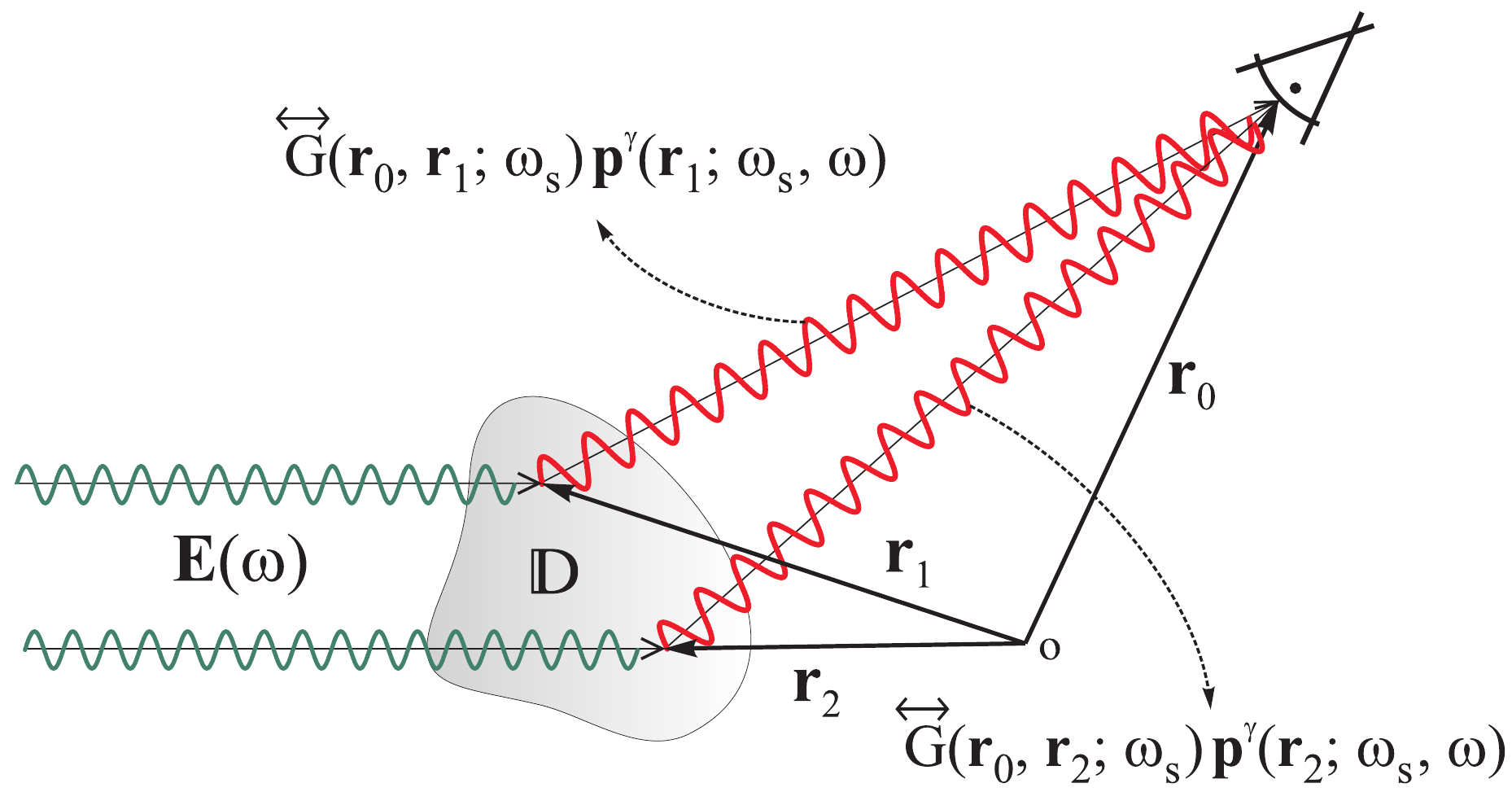}
\begin{narrowtext}
\caption[] {Illustration of two individual scattering paths associated with a scatterer $\mathbb{D}$ irradiated by the field ${\bf E}(\omega)$. On length scales $\left|{\bf r}_{1}\!-\!\!\;{\bf r}_{2}\right|$ smaller than the phonon correlation length $L_{\rm c}$ of a vibrational mode $\gamma$, the partial fields $\tensor{\rm G}({\bf r}_{0},{\bf r}_{1};\omega_{\rm s})\,{\bf p}^{\gamma}({\bf r}_{1};\omega_{\rm s},\omega)$ and $\tensor{\rm G}({\bf r}_{0},{\bf r}_{2};\omega_{\rm s})\,{\bf p}^{\gamma}({\bf r}_{2};\omega_{\rm s},\omega)$ add coherently at location ${\bf r}_{0}$ of the detector. On the other hand, for length-scales larger than $L_{\rm c}$ there is no phase-correlation between the scattering events and hence the partial fields at the detector add incoherently. \label{scatter}}
\end{narrowtext}
\end{figure}

For a linear scattering medium irradiated by a monochromatic incident beam of frequency $\omega$, the scattered field ${\bf E}^{\rm s}$ of frequency $\omega_{\rm s}$ that reaches the detector at the position ${\bf r}_{0}$ is described by the \emph{integral equation of potential scattering} of the form~\cite{wolfbook}\\
\begin{equation}
{\bf E}^{\rm s}({\bf r}_{0};\omega_{\rm s})\,=\frac{\omega_{\rm s}^2}{\varepsilon_{0}c^{2}}\,\int_{\mathbb{D}}\,d^{\,3}{\bf r}\,\tensor{\rm G}({\bf r}_{0},{\bf r};\omega_{\rm s})\,\mu({\bf r};\omega)\,{\bf E}({\bf r};\omega)\,,
\label{scattered}
\end{equation}\\
where $\varepsilon_{0}$ and $c$ are the free space permittivity and speed of light, respectively, $\mu({\bf r};\omega)$ is the scattering potential at a position ${\bf r}$ in the scattering domain $\mathbb{D}$, ${\bf E}({\bf r};\omega)$ is the total exciting field at ${\bf r}$, and $\tensor{\rm G}({\bf r}_{0},{\bf r};\omega_{\rm s})$ is the outgoing Green function which accounts for the whole system, including the scattering and surrounding media. In the case of vibrational Raman scattering, the scattering potential is described by the second-rank polarizability tensor $\tensor{\alpha}^{\,\gamma}$, whose components are defined as\\
\begin{equation}
\alpha^{\,\gamma}_{i,j} ({\bf r};\omega_{\rm s},\omega) \,=\, \sum_{k=x,y,z}\frac{\partial \alpha_{i,j}({\bf r};\omega)}{\partial q_k}\:q_k \;\,,
\label{polariz}
\end{equation}\\
with $\alpha$ being the polarizability per unit area at frequency $\omega$, and ${\bf q}=(q_x,q_y,q_z)$ being the lattice displacement vector associated with a particular vibrational mode $\gamma$ with frequency $\left|\omega-\omega_{\rm s}\right|$. Notice that the product $\tensor{\alpha}^{\,\gamma}({\bf r};\omega_{\rm s},\omega){\bf E}({\bf r};\omega)$ defines the induced Raman dipole per unit area ${\bf p}^{\,\gamma}({\bf r};\omega_{\rm s},\omega)$, and in this case Eq.~(\ref{scattered}) assumes the form\\
\begin{equation}
{\bf E}^{\rm s}({\bf r}_{0};\omega_{\rm s})\,=\frac{\omega_{\rm s}^2}{\varepsilon_{0}c^{2}}\,\int_{\mathbb{\mathbb{D}}}\,d^{\,3}{\bf r}\,\tensor{\rm G}({\bf r}_{0},{\bf r};\omega_{\rm s})\,{\bf p}^{\gamma}({\bf r};\omega_{\rm s},\omega)\,,
\label{scattereddipole}
\end{equation}\\
with\\
\begin{equation}
{\bf p}^{\gamma}({\bf r};\omega_{\rm s},\omega)=\tensor{\alpha}^{\,\gamma}({\bf r};\omega_{\rm s},\omega){\bf E}({\bf r};\omega)\;.\label{ramandipole}
\end{equation}\\
By considering the field as a single realization in spectral domain, the detector renders a signal $S({\bf r}_{0};\omega_{\rm s})$ that is proportional to the ensemble average of the scattered field~(\ref{scattereddipole}):\\
\begin{eqnarray}
S({\bf r}_{0};\omega_{\rm s})&=&\left\langle {\bf E}^{\rm s\,\ast}({\bf r}_{0},\omega_{\rm s})\cdot{\bf E}^{\rm s}({\bf r}_{0},\omega_{\rm s})\right\rangle\, \label{ensamble}  \\ [1ex]
&=&\frac{\omega_{\rm s}^2}{\varepsilon_{0}c^{2}}\,\int_{\mathbb{D}}\,d^{\,3}{\bf r}_{1}\,\int_{\mathbb{D}}\,d^{\,3}{\bf r}_{2}\,\left\langle\tensor{\rm G}({\bf r}_{0},{\bf r}_{1};\omega_{\rm s})\,{\bf p}^{\gamma}({\bf r}_{1};\omega_{\rm s},\omega)\cdot\tensor{\rm G}({\bf r}_{0},{\bf r}_{2};\omega_{\rm s})\,{\bf p}^{\gamma}({\bf r}_{2};\omega_{\rm s},\omega)\right\rangle\,, \nonumber
\end{eqnarray}\\
Figure~\ref{scatter} illustrates two individual scattering events, where the scattering domain $\mathbb{D}$ is irradiated by the field ${\bf E}(\omega)$. On length scales $\left|{\bf r}_{1}\!-\!\!\;{\bf r}_{2}\right|$ smaller than the phonon correlation length $L_{\rm c}$, the partial fields $\tensor{\rm G}({\bf r}_{0},{\bf r}_{1};\omega_{\rm s})\,{\bf p}^{\gamma}({\bf r}_{1};\omega_{\rm s},\omega)$ and $\tensor{\rm G}({\bf r}_{0},{\bf r}_{2};\omega_{\rm s})\,{\bf p}^{\gamma}({\bf r}_{2};\omega_{\rm s},\omega)$ add coherently at the detector. On the other hand, for length-scales larger than $L_{\rm c}$ there is no phase-correlation between the dipoles and hence the partial fields at the detector add incoherently. For experiments using a coherent exciting field (laser source), the signal becomes [(\ref{ramandipole}),(\ref{ensamble})]\\
\begin{eqnarray}
S({\bf r}_{0},\omega_{\rm s}) \!&=&\! \frac{\omega_{\rm s}^{4}}{\varepsilon_0^2 c^{4}}\,
 \sum_{l,m,n}\,\sum_{i,j}\;
\int_{\mathbb{D}}\,d^{3}{\bf r}_{2}\,\int_{\mathbb{D}}\,d^{3}{\bf r}_{1}\;
\big\langle \alpha^{\gamma\,\ast}_{mi}({\bf r}_{1},\omega_{\rm s})\,\alpha^{\gamma}_{nj}({\bf r}_{2},\omega_{\rm s})\big\rangle\,  \label{ramanintensity2}
\\[1ex]
&&\hspace{5em}\times\; {\rm G}_{lm}^{\ast}({\bf r}_{0},{\bf r}_{1};\omega_{\rm s}) \,{\rm G}_{ln}({\bf r}_{0},{\bf r}_{2};\omega_{\rm s})\;E^{\,\ast}_i({\bf r}_{1},\omega)\,E_j({\bf r}_{2},\omega)\;,
\nonumber
\end{eqnarray}\\
with $l,m,n\in\{x,y,z\}$. Eq.~(\ref{ramanintensity2}) tell us that, for experiments carried out with an incident laser beam, the spatial coherence of the scattered signal is solely described by the correlation of the Raman polarizability tensor components.\\

Classical textbooks describing Raman scattering usually do not consider the spatial coherence of the scattered field (see, for example, Refs.~~\cite{cardonaII,loudonbook}). The correlation function $\langle\alpha^{\gamma\,\ast}_{mi}\,\alpha^\gamma_{nj}\rangle$ is typically assumed to be a Dirac delta function, for which the signal in (\ref{ramanintensity2}) turns into a simple integration over the scattering volume $V_{\mathbb{D}}$. The outgoing Green function accounts for the polarization direction of the scattered field (defined by the unit vector $\hat{\bf \epsilon}_{\rm s}$) in the presence of analyzers, and also for the solid angle $\Omega$ of the collection optics. Eq.~(\ref{ramanintensity2}) is than reduced to~\cite{cardonaII,loudonbook}\\
\begin{equation}
S({\bf r}_{0},\omega_{\rm s})\,\propto\,V_{\mathbb{D}}\,\Omega\,\frac{\omega_{\rm s}^{4}}{\varepsilon_0^2 c^{4}}\,\left|\hat{\bf \epsilon}_{\rm s}\cdot\tensor{\alpha}^{\,\gamma}(\omega_{\rm s},\omega){\bf E}(\omega)\right|^{2}\,.
\label{classic}
\end{equation}\\
Indeed, spatial correlations associated with vibrational states can be neglected in usual Raman scattering experiments performed in the far-field regime, since the correlation length $L_{\rm c}$ of optical phonons in crystals is on the order of tens of nanometers, one order of magnitude shorter than the wavelength of visible light. In the case of Raman scattering of liquids and gases, this approximation is even better applied, since the correlation length associated with vibrational states of the molecules contained in these systems is in the range of a few nanometers, defined by thermal fluctuations.\\

The most important point here is that the analysis performed in Refs.~\cite{cardonaII,loudonbook} does not take into account the non-radiating near-field components in the light-matter interaction, and therefore Eq.~(\ref{classic}) hides important information related to the spatial correlation on length-scales smaller than $\lambda/2$. To account for spatial coherence in near-field Raman scattering, we will consider Gaussian correlations of the form\\
\begin{equation}
\big\langle \alpha^{\gamma\,\ast}_{mi}({\bf r}_{1},\omega_{\rm s})\,\alpha^{\gamma}_{nj}({\bf r}_{2},\omega_{\rm s})\big\rangle\ =
\tilde{\alpha}^{\gamma\,\ast}_{mi}({\bf r}_{1},\omega_{\rm s})\,
\tilde{\alpha}^{\gamma}_{nj}({\bf r}_{2},\omega_{\rm s})\;
\frac{{\rm e}^{-\left(\left|{\bf r}_{1}\!-\!\!\;{\bf r}_{2}\right|^2/ L_{\rm c}^2\right)}}{\pi\,L^2_{\rm c}}\;.\qquad
\label{correl}
\end{equation}\\
The last term turns into a spatial delta function in the limit $L_{\rm c}\rightarrow 0$, and into a constant term for $L_{\rm c}\rightarrow \infty$.\\

\begin{figure}
\includegraphics[scale=0.6]{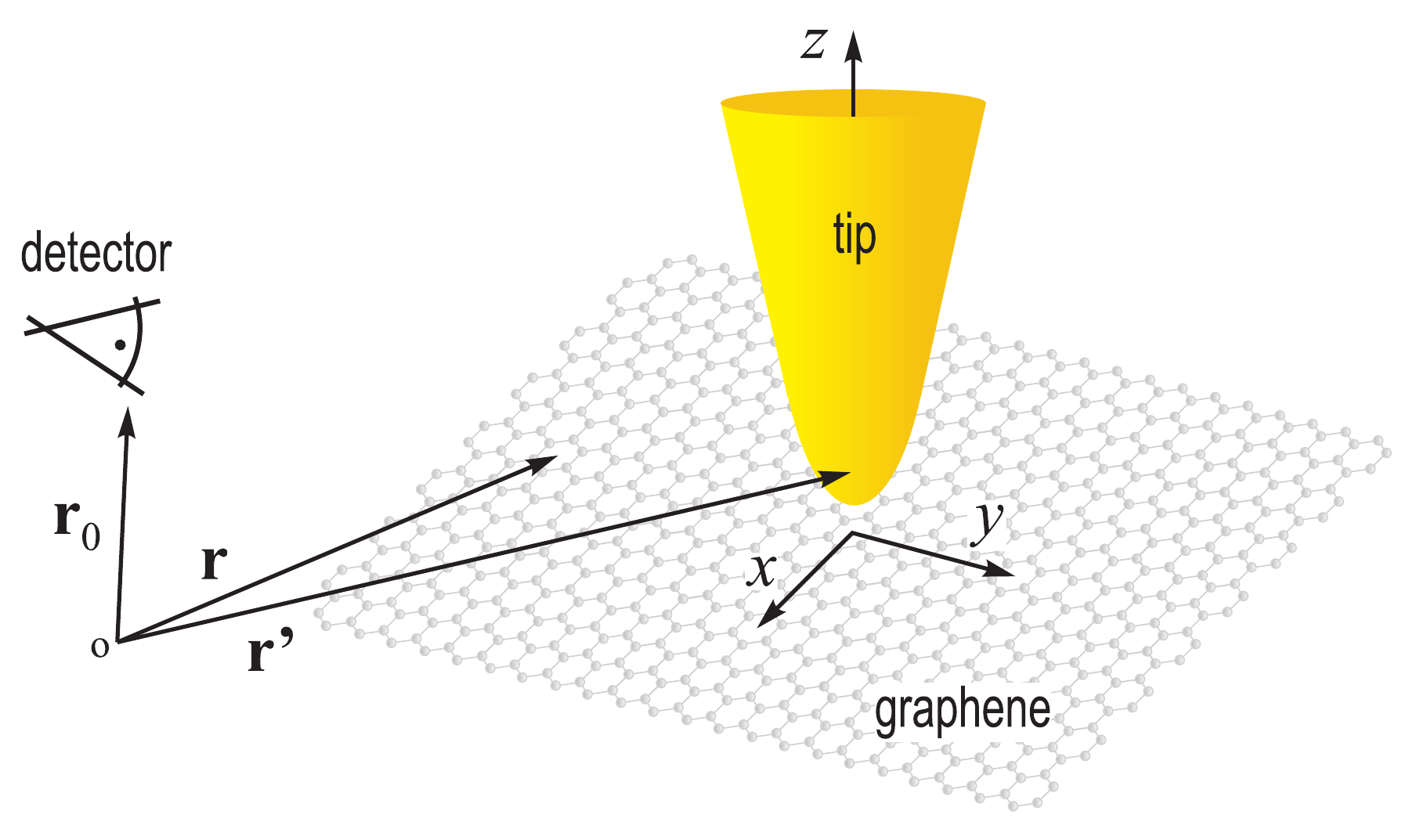}
\begin{narrowtext}
\caption[] {Illustration of spatially resolved near-field Raman scattering of a graphene sample. The electric field ${\bf E}$ confined to the apex of a laser-irradiated gold tip interacts locally with the graphene lattice characterized by the Raman polarizability $\tensor{\alpha}^{\,\gamma}$, where $\gamma$ denotes a specific phonon mode ($\gamma\in[{\rm G,\,D,\,G'}])$.  \label{schematic}}
\end{narrowtext}
\end{figure}

In the following we will consider monolayer graphene as our material system. The well-defined Raman modes of graphene provide an excellent model system for our theory, since they involve distinct symmetries and dimensionalities. The three main features present in the Raman spectrum of graphene are~\cite{adobook,ferrari2013}: (i) the one-phonon first-order allowed G band ($\sim$\,1580\,cm$^{-1}$) originating from the double-degenerate bond-stretching mode with E$_{\rm 2g}$ symmetry, occurring at the center of the Brillouin zone ($\Gamma$ point) where the transverse optical (TO) and longitudinal optical (LO) phonon branches touch each other; (ii) the disorder-induced D band ($\sim$\,1350\,cm$^{-1}$) originating from totally symmetric (A$_{1}$) TO phonons occurring near the edges (K and K$^{\prime}$ points) of the first Brillouin zone, activated by structural defects that provide momentum conservation in a double-resonance process; (iii) the two-phonon G$^{\prime}$ band (also called 2D) centered at $\sim$\,2700\,cm$^{-1}$, generated by triple-resonance processes in monolayer graphene, and related to the same phonon as the D band, although it does not require the presence of defects for its activation. While the G and G$^{\prime}$ bands are allowed over the entire graphene lattice, the defect-induced D band is strongly localized near the graphene edges, which gives it a one-dimensional character~\cite{casiraghi09b,lucchese10,beams11a,su13}. For this reason, we refer to G and G$^{\prime}$ as 2-D (two-dimensional), whereas D is denoted as a 1-D (one-dimensional) mode.  The corresponding Raman polarizability tensors associated with these Raman modes can be represented as~\cite{loudon64}\\
\begin{eqnarray}\label{alpharaman}
\tensor{\alpha}^{\,\rm G}({\rm E_{2g1}})=\alpha^{\rm G}
\left[
\begin{array}{rr}
1 & 0  \\
0 &-1
\end{array}
\right],\;\;\;\;
\tensor{\alpha}^{\,\rm G}({\rm E_{\rm 2g2}})=\alpha^{\rm G}
\left[
\begin{array}{rr}
0 & \;\;1  \\
1 & 0
\end{array}
\right],\;\;\;\;
\tensor{\alpha}^{\,\rm D,G^{\prime}}({\rm A_{1}})=\alpha^{\rm D,G^{\prime}}
\left[
\begin{array}{rr}
1 & \;\;0 \\
0 & 1
\end{array}
\right]\,,
\end{eqnarray}\\
where E$_{\rm 2g_{1}}$ and E$_{\rm 2g_{2}}$ are the two components of the double-degenerate E$_{\rm 2g}$ mode giving rise to the G band. We have omitted the $iz$ and $zj$ components in (\ref{alpharaman}), because they are null.\\

Figure~\ref{schematic} illustrates a near-field Raman experiment on a graphene sample, where ${\bf r'}=(0,0,z)$ denotes the position of the near-field probe,  ${\bf r}_{0}$ the location of the detector, and ${\bf r}=(x,y,0)$ is a point on the sample plane. The insertion of the Gaussian correlation function (\ref{correl}) into the expression for the signal (\ref{ramanintensity2}) yields\\
\begin{eqnarray}
S({\bf r}_{0},\omega_{\rm s}) &=&\frac{\omega_{\rm s}^{4}}{\varepsilon_0^2 c^{4}}\,
 \sum_{l,m,n}\sum_{i,j}\;
\int\!\!\!\int^{+\infty}_{-\infty} \!\!\!\!\! dx_{2} \,dy_{2}\; \,{\rm G}_{ln}({\bf r}_{0},x_{2},y_{2};\omega_{\rm s})\tilde{\alpha}^{\gamma}_{nj}\,E_j(x_{2},y_{2},\omega) \label{ramanintensity4}
\\[1ex]
&&\hspace{-1em}\times\;
\int\!\!\!\int^{+\infty}_{-\infty} \!\!\!\!\! dx_{1} \,dy_{1}\; \,
\frac{{\rm e}^{-[(x_{1}\;-\;x_{2})^2+\:\!(y_{1}\;-\;y_{2})^2] / L_{\rm c}^2}}{\pi\,L^2_{\rm c}}\;  {\rm G}_{lm}^{\ast}({\bf r}_{0},x_{1},y_{1};\omega_{\rm s}) \,\tilde{\alpha}^{\gamma\,\ast}_{mi}\,E^{\ast}_i(x_{1},y_{1},\omega)\;,
\nonumber
\end{eqnarray}\\
where $l\in\{x,y,z\}$ and $m,n;i,j\in\{x,y\}$.\\

To simplify notation, we will suppress the frequencies $\omega$ and $\omega_{\rm s}$ in the arguments of the different functions. We  introduce the Fourier transform of the correlation function\\
\begin{eqnarray}
\hspace{-1em} \frac{1}{4\pi^2}
\int\!\!\!\int^{+\infty}_{-\infty} \!\!\!\!\! \!dx_{2} \,dy_{2}\;
\big\langle \alpha^{\gamma\,\ast}_{mi}(x_{1},y_{1})\, \alpha^\gamma_{nj}(x_{2},y_{2})\big\rangle \;{\rm e}^{-i(k_{x} x_{2} + k_{y} y_{2})}\\
\;=\;   \frac{\tilde\alpha^{\gamma\,\ast}_{mi}\,
\tilde \alpha^\gamma_{nj}}{4\pi^2}\;
{\rm e}^{-i( k_x x_{1} + k_y y_{1}) - (k_x^2+k_y^2) L_{\rm c}^2/4}\;,\nonumber
\label{correlation2}
\end{eqnarray}\\
with $\tilde\alpha^\gamma_{ij}$  defined by the elements of the  Raman polarizability tensors (\ref{alpharaman}). The inverse transform of (\ref{correlation2}) is given as\\
\begin{eqnarray}
\big\langle \alpha^{\gamma\,\ast}_{mi}(x_{1},y_{1})\, \alpha^\gamma_{nj}(x_{2},y_{2})\big\rangle &=&
\int\!\!\!\int^{+\infty}_{-\infty} \!\!\!\!\! dk_x \:dk_y\;
 \frac{\tilde\alpha^{\gamma\,\ast}_{mi}\,
\tilde \alpha^\gamma_{nj}}{4\pi^2}\;
{\rm e}^{-(k_x^2+k_y^2) L_{\rm c}^2/4} \;{\rm e}^{i(k_x [x_{2}-x_{1}] + k_y [y_{2}-y_{1}])}\, . \;\;
\label{invtrf}
\end{eqnarray}\\
Likewise, we introduce the Fourier transform\\
\begin{eqnarray}
\hat{F}_{lnj}(k_x,k_y) &=& \frac{1}{4\pi^2}
\int\!\!\!\int^{+\infty}_{-\infty} \!\!\!\!\! dx_2 \,dy_2\;
{\rm G}_{ln}({\bf r}_{0},x_2,y_2) E_j(x_2,y_2)
 \;{\rm e}^{-i(k_x x_2 + k_y y_2)} \; ,
\label{ftransf01}
\end{eqnarray}\\
with the corresponding inverse transform
\begin{eqnarray}
{\rm G}_{ln}({\bf r}_{0},x_2,y_2) E_j(x_2,y_2)  &=&
\int\!\!\!\int^{+\infty}_{-\infty} \!\!\!\!\! dk_x \,dk_y\;
 \hat{F}_{lnj}(k_x,k_y)
 \;{\rm e}^{i(k_x x_2 + k_y y_2)} \; .
\label{ftransf01b}
\end{eqnarray}\\
Using these Fourier representations we can express the second integral in (\ref{ramanintensity4}) as
\begin{eqnarray}
&&\int\!\!\!\int^{+\infty}_{-\infty} \!\!\!\!\! dx_1 \,dy_1\; \,
\big\langle \alpha^{\gamma\,\ast}_{mi}(x_1,y_1)\, \alpha^\gamma_{nj}(x_2,y_2)\big\rangle\;  {\rm G}_{lm}^{\ast}({\bf r}_{0},x_1,y_1) \,E^{\ast}_i(x_1,y_1)
 \\[1ex]
&&\qquad =\;\frac{\tilde\alpha^{\gamma\,\ast}_{mi}\,
\tilde \alpha^\gamma_{nj}}{4\pi^2}\;\int\!\!\!\int^{+\infty}_{-\infty} \!\!\!\!\! dk_x \,dk_y\;  \hat{F}^{\ast}_{lmi}(k_x,k_y)
\int\!\!\!\int^{+\infty}_{-\infty} \!\!\!\!\! dk'_x \,dk'_y\;{\rm e}^{-({k'_x}^{\!2}+{k'_y}^{\!2}) L_{\rm c}^2/4}  \;{\rm e}^{i k'_x x_2 + i k'_y y_2} \, \nonumber \\
&& \hspace{6em} \times \int\!\!\!\int^{+\infty}_{-\infty} \!\!\!\!\! dx_1 \,dy_1\; \,
{\rm e}^{-i x_1 [k_x+k_x'] - i y_1 [k_y+k_y']}\,  \;\;
\nonumber \\[1ex]
&&\qquad=\;\tilde\alpha^{\gamma\,\ast}_{mi}\,
\tilde \alpha^\gamma_{nj} \int\!\!\!\int^{+\infty}_{-\infty} \!\!\!\!\! dk_x \,dk_y\;  \hat{F}^{\ast}_{lmi}(k_x,k_y)
\;{\rm e}^{-({k_x}^{\!2}+{k_y}^{\!2}) L_{\rm c}^2/4}  \;{\rm e}^{-i k_x x_2 - i k_y y_2} \; ,\nonumber
 \end{eqnarray}\\
where we used $\int\!e^{\,i (xy)}dx = 2\pi\delta(y)$. The signal (\ref{ramanintensity4}) can now be calculated as\\
\begin{eqnarray}
S({\bf r}_{0}) &=& \frac{\omega_{\rm s}^{4}}{\varepsilon_0^2 c^{4}}\,
 \sum_{l,m,n}\,\sum_{i,j}\;
\tilde\alpha^{\gamma\,\ast}_{mi}\,
\tilde \alpha^\gamma_{nj}
 \int\!\!\!\int^{+\infty}_{-\infty} \!\!\!\!\! dk_x \,dk_y\;  \hat{F}^{\ast}_{lmi}(k_x,k_y)\;
\;{\rm e}^{-({k_x}^{\!2}+{k_y}^{\!2}) L_{\rm c}^2/4}  \label{ramanintensity4mn}  \\
&&\hspace{0em} \times \int\!\!\!\int^{+\infty}_{-\infty} \!\!\!\!\! dk'_x \,dk'_y\;
 \hat{F}_{lnj}(k'_x,k'_y)
 \int\!\!\!\int^{+\infty}_{-\infty} \!\!\!\!\! dx_2 \,dy_2\; \, {\rm e}^{-i x_2 [k_x-k_x'] -i y_2 [k_y-k_y']}\, \nonumber \\[2ex]
 &=& 4\pi^2 \frac{\omega_{\rm s}^{4}}{\varepsilon_0^2 c^{4}}\,
 \sum_{l,m,n}\,\sum_{i,j}\;
\tilde\alpha^{\gamma\,\ast}_{mi}\,
\tilde \alpha^\gamma_{nj}
 \int\!\!\!\int^{+\infty}_{-\infty} \!\!\!\!\! dk_x \,dk_y\;  \hat{F}^{\ast}_{lmi}(k_x,k_y)\,  \hat{F}_{lnj}(k_x,k_y)\;
\;{\rm e}^{-({k_x}^{\!2}+{k_y}^{\!2}) L_{\rm c}^2/4} \, ,
\nonumber
 \end{eqnarray}\\
which, provided that the Fourier transform (\ref{ftransf01}) can be calculated, is considerably more convenient than the fourfold integral in (\ref{ramanintensity4}). In the fully coherent limit ($L_{\rm c}\rightarrow \infty$), there are no statistical variations between points ${\bf r}_{1}$ and ${\bf r}_{2}$, whereas in the incoherent case ($L_{\rm c}\rightarrow 0$), the response at  ${\bf r}_{1}$ and ${\bf r}_{2}$ is completely uncorrelated and the correlation function reduces to a Dirac delta distribution. In these limiting cases we find\\
\begin{eqnarray}
\hspace{-0em}\lim_{L_{\rm c}\rightarrow \infty} S({\bf r}_{0}) =  16 \pi^4\:\!\frac{\omega_{\rm s}^{4}}{\varepsilon_0^2 c^{4}}\:
 \sum_{l,m,n}\,\sum_{i,j}
 \alpha^{\gamma\,\ast}_{mi} \alpha^\gamma_{nj}
\hat{F}^{\ast}_{lmi}(0,0)\,  \hat{F}_{lnj}(0,0)\; \label{cohcase} \\
&&\hspace{-21em}=
\frac{\omega_{\rm s}^{4}}{\varepsilon_0^2 c^{4}}
 \sum_{l,m,n}\,\sum_{i,j}\!
\alpha^{\gamma\,\ast}_{mi} \alpha^\gamma_{nj}\int\!\!\!\int^{+\infty}_{-\infty} \!\!\!\!\!\! dx_1 \,dy_1\;
{\rm G}^{\ast}_{lm}({\bf r}_{0},x_1,y_1) E^{\ast}_i(x_1,y_1) \nonumber \\
&&\hspace{-12em}\times\int\!\!\!\int^{+\infty}_{-\infty} \!\!\!\!\!\! dx_2 \,dy_2\;
{\rm G}_{ln}({\bf r}_{0},x_2,y_2) E_j(x_2,y_2)\;, \nonumber
\nonumber
 \end{eqnarray}
\begin{eqnarray}
\lim_{L_{\rm c}\rightarrow 0} S({\bf r}_{0}) &=&
4\pi^2 \frac{\omega_{\rm s}^{4}}{\varepsilon_0^2 c^{4}}
 \sum_{l,m,n}\,\sum_{i,j}
\tilde\alpha^{\gamma\,\ast}_{mi}
\tilde \alpha^\gamma_{nj}
 \int\!\!\!\int^{+\infty}_{-\infty} \!\!\!\!\! dk_x \,dk_y\;  \hat{F}^{\ast}_{lmi}(k_x,k_y)\,  \hat{F}_{lnj}(k_x,k_y) \;
\label{cohcase} \\
&=&
\frac{\omega_{\rm s}^{4}}{\varepsilon_0^2 c^{4}}\,
 \sum_{l,m,n}\,\sum_{i,j}\;
\tilde\alpha^{\gamma\,\ast}_{mi}\,
\tilde \alpha^\gamma_{nj}
\int\!\!\!\int^{+\infty}_{-\infty} \!\!\!\!\! dx \,dy\;
{\rm G}^{\ast}_{lm}({\bf r}_{0},x,y) E^{\ast}_i(x,y) \,
{\rm G}_{ln}({\bf r}_{0},x,y) E_j(x,y) \; .
\nonumber
 \end{eqnarray}\\
where $\alpha^{\gamma\,\ast}_{mi} \alpha^\gamma_{nj} = \tilde\alpha^{\gamma\,\ast}_{mi}\tilde \alpha^\gamma_{nj}\,/\,\pi L_{\rm c}^2$. These two limits were studied in Refs.~\cite{cancado2009,maximiano2011} for 1-D and 2-D systems, respectively. Here we discuss the more realistic case, where the correlation length $L_{\rm c}$ has a finite value.\\

In order to evaluate the signal in (\ref{ramanintensity4}), we need to know the exciting field ${\bf E}$ at the location ${\bf r}=(x,y,0)$ on the graphene plane. This field corresponds to the superposition of the incident laser field ${\bf E}_{0}$ and the localized field generated by the gold tip acting as an optical antenna. It can be represented in terms of a volume integral equation as~\cite{nanobook}\\
\begin{equation}
{\bf E}({\bf r},\omega) \;=\;  {\bf E}_{0}({\bf r},\omega) \,+\, \frac{\omega^2}{c^2}\! \int \!\!d^3{\bf r}^{\prime\prime}\,\tensor{\rm G}({\bf r},{\bf r}^{\prime\prime},\omega) \left[\varepsilon({\bf r}^{\prime\prime})\!-\!1\right] {\bf E}({\bf r}^{\prime\prime},\omega) \, ,
\label{viq}
\end{equation}\\
where $\varepsilon({\bf r}^{\prime\prime}\!\!,\omega)$ is the spatial distribution of the dispersive dielectric constant and $\tensor{\rm G}$ is the Green function of the reference system, which includes the sample, the supporting surface, and the tip. In principle, the field (\ref{viq}) can be substituted into the expression for the signal in (\ref{ramanintensity4}), and the signal can be computed numerically. To reduce the numerical complexity we describe the  tip by an anisotropic polarizability~\cite{nanobook}\\
\begin{eqnarray}
\tensor\alpha_{\rm tip}({\bf r}^{\prime}) =
\left[
\begin{array}{ccc}
\alpha_{\perp} & 0 & 0 \\
0 & \alpha_{\perp} & 0 \\
0 & 0 & \alpha_{\parallel}
\end{array}
\right] \;,
\label{tip}
\end{eqnarray}\\
with the tip axis coinciding with the $z$-direction. $\alpha_{\perp}$ and $\alpha_{\parallel}$ denote the transverse and longitudinal polarizabilities defined as~\cite{nanobook}\\
\begin{equation}
\alpha_{\perp}(\omega)=4\pi\,\varepsilon_{\rm 0}\,r_{\rm tip}^{3} \frac{\varepsilon(\omega)-1}{\varepsilon(\omega)+2}\,,
\label{tip02}
\end{equation}
and
\begin{equation}
\alpha_{\parallel}(\omega)=2\pi\,\varepsilon_{\rm 0}\,r_{\rm tip}^{3}\,f_{\rm e}(\omega),
\label{tip03}
\end{equation}\\
where $\varepsilon$ denotes the dielectric constant of the tip, $r_{\rm tip}$ is the tip radius, and $f_{\rm e}$ is the complex field enhancement factor. Accordingly, a general field ${\bf E}$ interacting with the tip induces a dipole ${\bf p}_{\rm tip} = \tensor\alpha_{\rm tip} {\bf E}$ in the tip. The Green function in (\ref{ramanintensity4}) can then be written as\\
\begin{eqnarray}
\tensor{\rm G}({\bf r}_{0},{\bf r};\omega_{\rm s}) &=& \tensor{\rm G}^{\rm o}({\bf r}_{0},{\bf r};\omega_{\rm s}) \;+\;
\frac{\omega_{\rm s}^{2}}{\varepsilon_0 c^{2}}\,
\tensor{\rm G}^{\rm o}({\bf r}_{0},{\bf r};\omega_{\rm s})\,\tensor\alpha_{\rm tip}({\bf r}^{\prime};\omega_{\rm {\rm s}}) \,
\tensor{\rm G}^{\rm o}({\bf r}',{\bf r};\omega_{\rm s}) \;,
\label{gnfct2}
\end{eqnarray}\\
where the first term denotes the free-space propagation from a point ${\bf r}=(x,y,0)$ on the graphene sample to the observation point ${\bf r}_{\rm 0}$, and the second term corresponds to the interaction with the tip dipole at ${\bf r}'=(0,0,z)$, that is, free-space propagation from graphene to the tip and subsequent propagation from tip to the observation point. We apply the same model for the excitation field described by (\ref{viq}) and obtain\\
\begin{eqnarray}
{\bf E}({\bf r},\omega) &=&
{\bf E}_0({\bf r},\omega)\,+\,\frac{\omega^{2}}{\varepsilon_0 c^{2}} \:
\tensor{\rm G}^{\rm o}({\bf r},{\bf r}';\omega)\, \tensor{\alpha}_{\rm tip}(\omega)\,{\bf E}_0({\bf r}',\omega).
\label{excfield}
\end{eqnarray}\\

\begin{figure}
\includegraphics[scale=0.8]{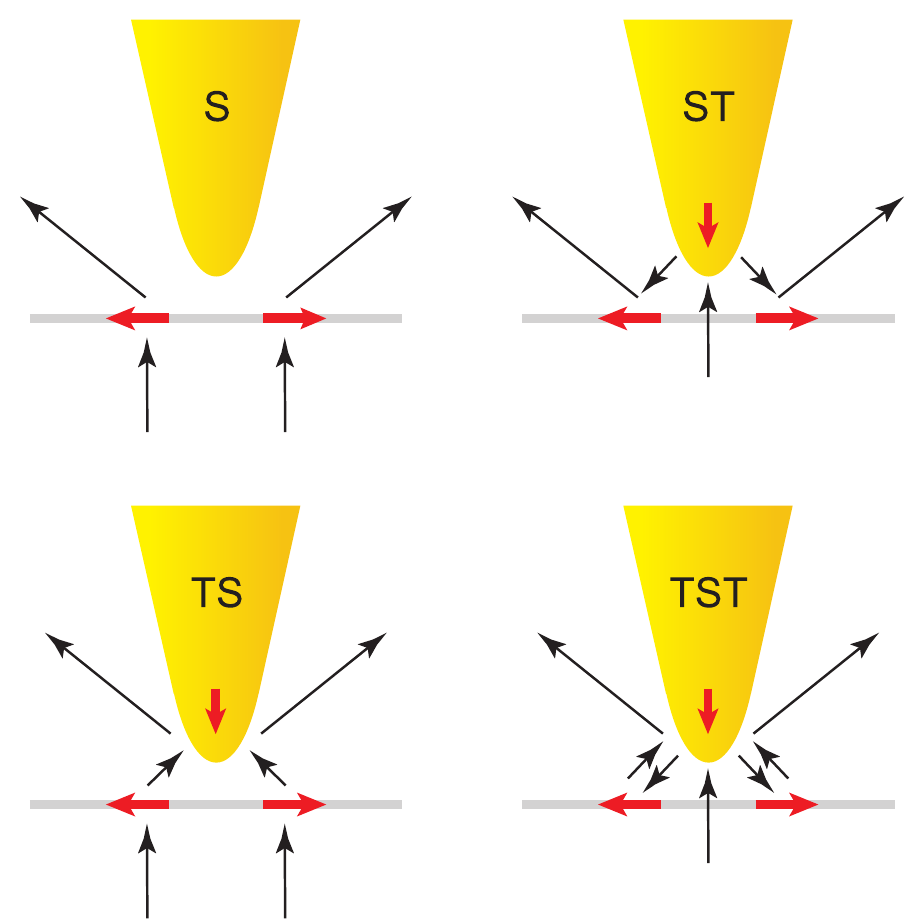}
\begin{narrowtext}
\caption[] {Interaction series in tip-enhanced Raman scattering (TERS). Red arrows represent induced dipoles, black arrows indicate electromagnetic interactions. \label{series}}
\end{narrowtext}
\end{figure}

Equations (\ref{gnfct2}) and (\ref{excfield}) can now be inserted into (\ref{ramanintensity4}) to calculate the signal $S({\bf r}_{\rm o},\omega_{\rm s})$ in terms of the free-space Green function and as a function of the tip position and excitation conditions. Combining (\ref{gnfct2}) and (\ref{excfield}) gives\\
\begin{eqnarray}
\hspace{4em}&&\hspace{-5em}{\rm G}_{ln}({\bf r}_{\rm o},x,y;\omega_{\rm s})\,\alpha_{nj}^{\gamma}(x,y;\omega)\,E_j(x,y,\omega) \;=\;  {\rm G}_{ln}^{\rm o}({\bf r}_{\rm o},x,y;\omega_{\rm s})\,\alpha_{nj}^{\gamma}(x,y;\omega)\,E_{0_j}(x,y,\omega) \;+\; \nonumber \\[0.5ex]
&&\frac{\omega_{\rm s}^{2}}{\varepsilon_0 c^{2}}\,
\left[\tensor{\rm G}^{\rm o}({\bf r}_{\rm o},z;\omega_{\rm s})\,\tensor\alpha_{\rm tip}(\omega_{\rm s})\,
\tensor{\rm G}^{\rm o}(z,x,y;\omega_{\rm s})\right]_{ln} \,\alpha_{nj}^{\gamma}(x,y;\omega)\,E_{0_j}(x,y,\omega) \;+\; \nonumber \\
&&\frac{\omega^{2}}{\varepsilon_0 c^{2}} \:{\rm G}_{ln}^{\rm o}({\bf r}_{\rm o},x,y;\omega_{\rm s})\,\alpha_{nj}^{\gamma}(x,y;\omega)\,\left[\tensor{\rm G}^{\rm o}(x,y,z;\omega)\, \tensor{\alpha}_{\rm tip}(\omega)\,{\bf E}_0(z,\omega) \right]_j
 \;+\;  \label{excfield3}  \\
&&\frac{\omega^{2}\:\!\omega_{\rm s}^{2}}{\varepsilon_0^2 c^{4}}\,
\left[\tensor{\rm G}^{\rm o}({\bf r}_{\rm o},z;\omega_{\rm s})\,\tensor\alpha_{\rm tip}(\omega_{\rm s})\,
\tensor{\rm G}^{\rm o}(z,x,y;\omega_{\rm s})\right]_{ln}\nonumber  \\
&&\;\times\;\alpha_{nj}^{\gamma}(x,y;\omega)\,
\left[\tensor{\rm G}^{\rm o}(x,y,z;\omega)\, \tensor{\alpha}_{\rm tip}(\omega)\,{\bf E}_0(z,\omega) \right]_j  \nonumber
\end{eqnarray}\\
where we explicitly used ${\bf r}'=(0,0,z)$ and ${\bf r}=(x,y,0)$. Since the polarizability along the tip axis is much larger than transverse to it ($|\alpha_{\parallel}|\gg|\alpha_{\perp}|$) we neglect  $\alpha_{\perp}$ in [(\ref{tip}),(\ref{excfield3})] and obtain\\
\begin{eqnarray}
\hspace{4em}&&\hspace{-5em}{\rm G}_{ln}({\bf r}_{\rm o},x,y;\omega_{\rm s})\,\alpha_{nj}^{\gamma}(x,y,\omega)\,E_j(x,y,\omega) \;= \nonumber\\
[1.5ex]
&&\hspace{-5em}{\rm G}_{ln}^{\rm o}({\bf r}_{\rm o},x,y;\omega_{\rm s})\,\alpha_{nj}^{\gamma}(x,y,\omega)\,E_{0\,j}(x,y,\omega) \;+\; \nonumber  \\
&&\hspace{-5em}\frac{\omega_{\rm s}^{2}}{\varepsilon_0 c^{2}}\,
{\rm G}_{lz}^{\rm o}({\bf r}_{\rm o},z^{\,\prime};\omega_{\rm s})\,\alpha_{\parallel}(\omega_{\rm s})\, \,{\rm G}_{zn} ^{\rm o}(z^{\,\prime},x,y;\omega_{\rm s})\,\alpha_{nj}^{\gamma}(x,y,\omega)\,E_{0\,j}(x,y,\omega) \;+\; \label{excfield4c} \\
&&\hspace{-5em}\frac{\omega^{2}}{\varepsilon_0 c^{2}} \,
{\rm G}_{ln}^{\rm o}({\bf r}_{\rm o},x,y;\omega_{\rm s})\,\alpha_{nj}^{\gamma}(x,y,\omega)\,{\rm G}_{jz}^{\rm o}(x,y,z^{\,\prime};\omega)\, \alpha_{\parallel}(\omega)\,E_{0\,z}(z^{\,\prime},\omega)
 \;+\; \nonumber   \\
&&\hspace{-5em}\frac{\omega^{2}\omega_{\rm s}^2}{\varepsilon_0^2 c^{4}}\,
\;\!{\rm G}_{lz}^{\rm o}({\bf r}_{\rm o},z^{\,\prime};\omega_{\rm s})\,\alpha_{\parallel}(\omega_{\rm s})\,\,{\rm G}_{zn}^{\rm o}(z^{\,\prime},x,y;\omega_{\rm s})
\,\alpha_{nj}^{\gamma}(x,y,\omega)\,{\rm G}_{jz}^{\rm o}(x,y,z^{\,\prime};\omega)\,\alpha_{\parallel}(\omega)\, E_{0\,z}(z^{\,\prime},\omega)\,.   \nonumber
\end{eqnarray}\\
The first term in (\ref{excfield4c}) is the interaction of the incident field with the sample ({\bf S}), the second accounts for the incident field that scatters at the sample and then at the tip ({\bf TS}), the third is the interaction with the tip and then with the sample ({\bf ST}), and the last term is the interaction  with the tip, then the sample and  the tip again ({\bf TST}). In other words, (\ref{excfield4c}) describes the following interaction series~\cite{carney2007}\\
\begin{equation}
{\bf S}+{\bf TS}+{\bf ST}+{\bf TST}\,,
\label{eqseries}
\end{equation}\\
which is illustrated in Figure~\ref{series}. In our scheme, the ${\bf S}$ term in (\ref{excfield4c}) describes standard far-field Raman scattering (e.g. confocal Raman scattering) and the ${\bf TST}$ accounts for tip-enhanced Raman scattering, TERS. The terms ${\bf ST}$ and ${\bf TS}$ originate from the interference  between the near-field and the far-field.  We have suppressed higher-order interaction terms between the graphene sample and the tip because graphene interacts only weakly with light (3\% absorption). TERS relies on excitation fields that exhibit a strong polarization component along the tip axis. Such conditions can be provided, for example, by a focused  radially polarized laser beam incident along the tip axis~\cite{achim2003}. Because the $z$ component of the incident field is much larger than the in-plane $x,y$ components, the signal strength generated by the {\bf TS} component is roughly 10$\times$ weaker than by the {\bf ST} term. For this reason, we neglect the {\bf TS} term in the present study.\\

We have now all ingredients to evaluate the {\bf TST} and {\bf ST} components of the near-field Raman signal for 2-D and 1-D samples.\\

\textbf{TST/2-D:}\\

\begin{figure}
\includegraphics[scale=1.2]{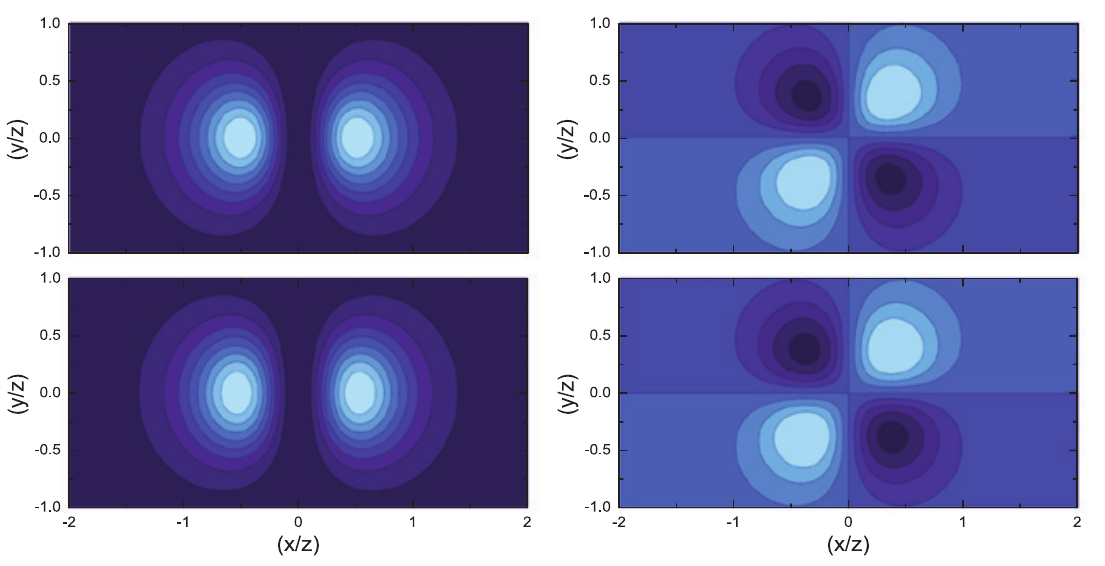}
\vspace{-1em}
\begin{narrowtext}
\caption[] {Comparison of original function (top) and approximation (bottom) according to (\ref{fttrf09}). The left panels correspond to $n=j$($=x$ in this case), while the right panels correspond to $n\neq j$. In both cases, the following fitting constants were used:  $a_0=0.74$, $b_0=4.0$, $c_0=0.08$, and $d_0=1.5$. \label{Fit_TST}}
\end{narrowtext}
\end{figure}

In the following, we proceed with the evaluation of the $S^{\rm TST}({\bf r}_{0})$ signal generated by the {\bf TST} component for the 2-D modes (G and G$^{\prime}$). We assume that the exciting field is homogeneous, that is $E_{0\,z}(z,\omega) =E_{0\,z}(\omega)$, which is fulfilled for tip-sample distances $z$ much smaller than $\lambda$. Considering the {\bf TST} term in (\ref{excfield4c}) we can express the Fourier transform of ${\rm G}_{ln}({\bf r}_{0},x,y;\omega_{\rm s}) E_j(x,y,\omega)$ [Eq.~(\ref{ftransf01})] as\\
\begin{eqnarray}
\hat{F}_{lnj}(k_x,k_y) &=& \frac{1}{4\pi^2} \,\frac{\omega^{2}\omega_{\rm s}^2}{\varepsilon_0^2 c^{4}}\,
\alpha_{\parallel}(\omega)\;\!\alpha_{\parallel}(\omega_{\rm s})\,{\rm G}_{lz}^{\rm o}({\bf r}_{0},z;\omega_{\rm s})\,E_{0\,z}(\omega) \label{ftransf09}
 \\[0ex]
 &&\hspace{0em} \times
\int\!\!\!\int^{+\infty}_{-\infty} \!\!\!\!\! dx \,dy
\,{\rm G}_{zn}^{\rm o}(x,y,z;\omega_{\rm s}) \,{\rm G}_{jz}^{\rm o}(x,y,z;\omega)\,
 {\rm e}^{-i(k_x x + k_y y)} \; .
\nonumber
\end{eqnarray}\\
Because of the short-range interaction between tip and sample ($R=|{\bf r}-{\bf r}^{\,\prime} |\ll \lambda,\lambda_{\rm s}$), we retain only the non-retarded near-field term in the Green functions ${\rm G}_{zn}^{\rm o}$ and ${\rm G}_{jz}^{\rm o}$, that is,\\
\begin{eqnarray}
{\rm G}_{zn}^{\rm o}(\omega_{\rm s}) \,=\,\!\frac{1}{4\pi k_{\rm s}^{2}} \frac{3 zn}{R^{\,5}}\;\;\;{\rm and}\;\;\; {\rm G}_{jz}^{\rm o}(\omega) \,=\,\!\frac{1}{4\pi k^2}\,\frac{3 jz}{R^{\,5}}   \;,
\label{ch3031}
\end{eqnarray}\\
where $k_{\rm s}=\omega_{\rm s}/c$, and $k=\omega/c$. The integral in (\ref{ftransf09}) now becomes\\
\begin{eqnarray}
&&\int\!\!\!\int^{+\infty}_{-\infty} \!\!\!\!\! dx \,dy
\,{\rm G}_{zn}^{\rm o}(x,y,z;\omega_{\rm s}) \,{\rm G}_{jz}^{\rm o}(x,y,z;\omega)\,
 {\rm e}^{-i(k_x x + k_y y)} \label{ftransf10} \\[0ex]
&& \hspace{4em} \;=\; \frac{9}{16\pi^2}\frac{c^{4}}{\omega_{\rm s}^{2}\omega^{2}}\:
  \;\int\!\!\!\int^{+\infty}_{-\infty} \!\!\!\!\! dx \,dy\; \frac{n j z^2}{(x^2+y^2+z^2)^5} \,
 {\rm e}^{-i(k_x x + k_y y)} \, . \;\;\;
 \nonumber
\end{eqnarray}\\
This integral has a complicated analytical form and hence it is convenient to approximate the integrand by a superposition of exponential functions\\
\begin{eqnarray}
\frac{(nj/z^2)}{z^6\,[(x/z)^2+(y/z)^2+1]^5} \;\approx \;  \frac{(nj/z^2)}{z^6}\,\left[ a_0\,{\rm e}^{-b_0 \;\!(x^2 + y^2)/z^2}\,+\,c_0\,{\rm e}^{-d_0 \;\!(x^2 + y^2)/z^2}\right] \, ,
\label{fttrf09}
\end{eqnarray}\\
where $a_0$, $b_0$, $c_0$, and $d_0$ are fitting constants. Figure~\ref{Fit_TST} shows a comparison of the original function and the approximation. The fitting constants of the latter are given in the caption. The agreement is very good, and the approximation is not expected to have any influence on the results of this study.\\

The Fourier transform (\ref{ftransf09}) is now calculated as\\
\begin{equation}
\hat{F}_{lnj}(k_x,k_y)\,=\,\frac{9}{256\,\pi^3 \,\varepsilon_0^2\, z^{4}}\,
\alpha_{\parallel}(\omega)\;\!\alpha_{\parallel}(\omega_{\rm s})\,{\rm G}_{lz}^{\rm o}({\bf r}_{0},z;\omega_{\rm s})\,E_{0\,z}(z,\omega)\, \hat{h}_{nj}(k_x,k_y;z)\;,\label{ftransf20}
\end{equation}\\
with\\
\begin{eqnarray}\label{function_h}
&\hat{h}_{nj}(k_x,k_y;z)=&\bigg[\frac{a_0 (2\delta_{nj} b_0 - k_n k_j z^2)}{b_0^3} {\rm e}^{- (k_x^2 + k_y^2) z^2\,/\, 4 b_0} \,\\[0ex]
&&\hspace{0em}+\,\frac{c_0 (2\delta_{nj} d_0 - k_n k_j z^2)}{d_0^3} {\rm e}^{- (k_x^2 + k_y^2) z^2\,/\, 4 d_0}\bigg]\, ,\;\;\;\; \nonumber
\end{eqnarray}\\
$\delta_{nj}$ being a Kronecker delta. To calculate the Raman signal (\ref{ramanintensity4mn}) for the {\bf TST} interaction term we define the expressions \\
\begin{eqnarray}
f_{mi,nj}(z,L_{\rm c})\;=\;  \int\!\!\!\int^{+\infty}_{-\infty} \!\!\!\!\! dk_x \,dk_y\;  \hat{h}^{\ast}_{mi}(k_x,k_y;z)\,  \hat{h}_{nj}(k_x,k_y;z)\;
\;{\rm e}^{-({k_x}^{\!2}+{k_y}^{\!2}) L_{\rm c}^2/4} \,, \label{fminj}
 \end{eqnarray}\\
which can be calculated analytically using the result in (\ref{function_h}). The functions $f_{mi,nj}$ have the properties\\
\begin{eqnarray}
f_{mi,nj}(z, L_{\rm c}\rightarrow\infty) = 1 / L_{\rm c}^2\,\,,\qquad f_{mi,nj}(z, L_{\rm c}\rightarrow 0) = 1/z^2 \; .
\end{eqnarray}\\
Inserting (\ref{fminj}) into (\ref{ramanintensity4mn}) yields\\
\begin{eqnarray}
S^{\rm TST}({\bf r}_{0}) &=& \frac{(3/8)^4}{4 \pi^4}\,\frac{\omega_{\rm s}^{4}}{\varepsilon_0^6 c^{4}z^{8}}\,
\sum_{l}\;
\left|\alpha_{\parallel}(\omega)\;\!\alpha_{\parallel}(\omega_{\rm s})\,{\rm G}_{lz}^{\rm o}({\bf r}_{0},z;\omega_{\rm s})\,E_{0\,z}(z,\omega)\right|^2 \times \label{reslt01} \\
&&  \hspace{5em}\;\; \left[\sum_{m,n}\,\sum_{i,j} \tilde\alpha^{\gamma\,\ast}_{mi}\,
\tilde \alpha^\gamma_{nj} \, f_{mi,nj}(z,L_{\rm c})\, \right] . \nonumber \;\;\;\;\;
 \end{eqnarray}\\
Thus, we find that for graphene with zero correlation length ($L_{\rm c}\rightarrow 0$) the signal decays as $z^{-10}$ from the graphene sample, whereas for graphene with infinite correlation length ($L_{\rm c}\rightarrow \infty$) it decays as $z^{-8}$, consistent with the theory described in Ref~\cite{maximiano2011}.\\

\begin{figure}
\includegraphics[scale=0.90]{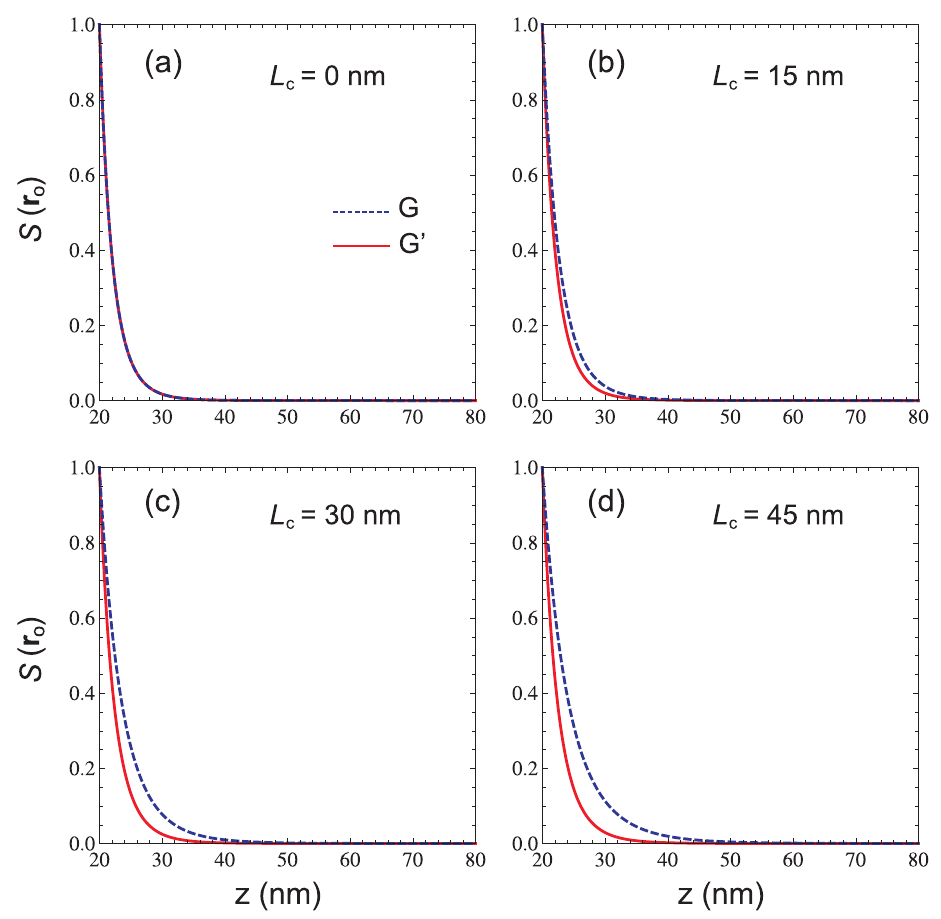}
\begin{narrowtext}
\caption[] { Distance dependence of the {\bf TST} signal for the G and G$^{\prime}$ bands. Panels (a-d) are evaluated for different correlation lengths $L_{\rm c}$: 0, 15, 30, and 45\,nm, respectively. In all cases we used $\tilde{f}_{\rm e}=3$. The signal $S({\bf r}_{0})$ is normalized to $1$ for $z_{\rm min}=20\,$nm, which corresponds to the closest tip-sample distance for $r_{\rm tip}=15\,$nm, considering that the minimal distance between the tip apex and the sample is $\sim\,5$\,nm. For $L_{\rm c}=0$ the curves for G and G$^{\prime}$ coincide. This case corresponds to an incoherent addition of the Raman response in different points of the graphene lattice. As $L_{\rm c}$ increases, interferences between neighboring lattice points give rise to a distance dependence that is different for the G and G$^{\prime}$ bands.\label{results_TST}}
\end{narrowtext}
\end{figure}

For the modes with E$_{\rm 2g1}$ and A$_{1}$ symmetries, for which the only non-null terms in the respective Raman tensors are $\tilde\alpha_{xx}^{\gamma}$ and $\tilde\alpha_{yy}^{\gamma}$ [see Eq.~(\ref{alpharaman})], the signal $S({\bf r}_{0})$ is given as:\\
\begin{eqnarray}
S^{\rm TST}({\bf r}_{0})({\rm E_{2g1},A_{1}}) &=& \frac{(3/8)^4}{4 \pi^4 \varepsilon_0^4\,z^8}\,\frac{\omega_{\rm s}^{4}}{\varepsilon_0^2 c^{4}}\,
\sum_{l}\;
\left|\alpha_{\parallel}(\omega)\;\!\alpha_{\parallel}(\omega_{\rm s})\,{\rm G}_{lz}^{\rm o}({\bf r}_{0},z;\omega_{\rm s})\,E_{0\,z}(z,\omega)\right|^2 \times \label{reslt01} \\
&&\Big(\left[\tilde\alpha^{\gamma\,\ast}_{xx}\,
\tilde \alpha^{\gamma}_{xx} +  \tilde\alpha^{\gamma\,\ast}_{yy}\,
\tilde \alpha^{\gamma}_{yy} \right] \, f_{xx,xx}(z,L_{\rm c})   \,+\, 2\,{\rm Re}\left [\tilde\alpha^{\gamma\,\ast}_{xx}\,
\tilde \alpha^{\gamma}_{yy}\,
f_{xx,yy}(z,L_{\rm c}) \right]\,     \Big)\,, \nonumber \;\;\;\;\;
\end{eqnarray}\\
where we have used the properties $f_{xx,xx}(z,L_{\rm c})=f_{yy,yy}(z,L_{\rm c})$, and $f_{xx,yy}(z,L_{\rm c})=f_{yy,xx}(z,L_{\rm c})$. On the other hand, modes belonging to the ${\rm E_{2g2}}$ irreducible representation, for which $\tilde\alpha_{xy}^{\gamma}=\tilde\alpha_{yx}^{\gamma}\neq0$ [see Eq.~(\ref{alpharaman})], we have\\
\begin{eqnarray}
S^{\rm TST}({\bf r}_{0})({\rm E_{2g2}}) &=& \frac{(3/8)^4}{4 \pi^4 \varepsilon_0^4\,z^8}\,\frac{\omega_{\rm s}^{4}}{\varepsilon_0^2 c^{4}}\,
\sum_{l}\;
\left|\alpha_{\parallel}(\omega)\;\!\alpha_{\parallel}(\omega_{\rm s})\,{\rm G}_{lz}^{\rm o}({\bf r}_{0},z;\omega_{\rm s})\,E_{0\,z}(z,\omega)\right|^2 \times \label{reslt02} \\
&&  \hspace{5em}\;\;4\,{\rm Re}\left [\tilde\alpha^{\gamma\,\ast}_{xy}\,
\tilde \alpha^{\gamma}_{yx}\,
f_{xy,yx}(z,L_{\rm c}) \right]\,, \nonumber \;\;\;\;\;
\end{eqnarray}\\
where we have used the fact that $f_{xy,yx}(z,L_{\rm c})=f_{yx,xy}(z,L_{\rm c})=f_{xy,xy}(z,L_{\rm c})=f_{yx,yx}(z,L_{\rm c})$.\\

Next, we rewrite the tip polarizability component $\alpha_{\parallel}(\omega)$ in terms of the complex field enhancement factor $f_{\rm e}(\omega)$ and the tip radius $r_{\rm tip}$ [Eq.~(\ref{tip03})], and insert the values of the polarizability tensors (\ref{alpharaman}) in Eqs.~(\ref{reslt01}) and (\ref{reslt02}). For the G band we find\\
\begin{eqnarray}
S^{\rm TST}_{\rm G}({\bf r}_{0}) &=&\frac{81}{512}\,\frac{\omega_{\rm s}^{4}}{\varepsilon_0^2 c^{4}}\frac{r_{\rm tip}^{12}\,\tilde{f}_{\rm e}(\omega)^{4}}{z^8}\,
\sum_{l}\;\left|{\rm G}_{lz}^{\rm o}({\bf r}_{0},z;\omega_{\rm s})\,E_{0\,z}(z,\omega)\: \tilde\alpha^{\rm G}(\omega_{\rm s};\omega)
 \right|^2  \quad \label{reslt03} \\[-1ex]
&&  \hspace{4em}\;\; \times \left[ f_{xx,xx}(z,L_{\rm c})   \,-\, f_{xx,yy}(z,L_{\rm c}) \,+\,2f_{xy,yx}(z,L_{\rm c})\right]\, , \nonumber \;\;\;\;\;
 \end{eqnarray}\\
where we have considered $\tilde{f}_{\rm e}(\omega)\approx\tilde{f}_{\rm e}(\omega_{s})$, $\tilde{f}_{\rm e}$ being the real part of $f_{\rm e}$. For the Raman G$^{\prime}$ band we obtain\\
\begin{eqnarray}
S^{\rm TST}_{\rm G^{\prime}}({\bf r}_{0}) &=&\frac{81}{512}\,\frac{\omega_{\rm s}^{4}}{\varepsilon_0^2 c^{4}}\frac{r_{\rm tip}^{12}\,\tilde{f}_{\rm e}(\omega)^{4}}{z^8}\,
\sum_{l}\;
\left|{\rm G}_{lz}^{\rm o}({\bf r}_{0},z;\omega_{\rm s})\,E_{0\,z}(z,\omega)\: \tilde\alpha^{\rm G'}(\omega_{\rm s};\omega)
 \right|^2  \quad \label{reslt04} \\[-1ex]
&&  \hspace{4em}\;\; \times \left[ f_{xx,xx}(z,L_{\rm c})   \,+\, f_{xx,yy}(z,L_{\rm c})\right]\, . \nonumber \;\;\;\;\;
\end{eqnarray}\\
The dependence on the correlation length $L_{\rm c}$ and the distance $z$ is solely determined by the last terms in the expressions. The first terms  only account for the overall strength of the signal. They depend on the Raman cross section, the local field enhancement, the tip radius, and the detection conditions. In the following we will discuss the consequences of the different terms in (\ref{reslt03}) and (\ref{reslt04}).\\

In Figure~\ref{results_TST} we plot the distance dependence of the G and G$^{\prime}$ {\bf TST} signals [according to Eqs. (\ref{reslt03}) and (\ref{reslt04}), respectively] for different correlation lengths $L_{\rm c}$. For all cases, we set $\tilde{f}_{\rm e}(\omega)\,=\,3$ and $r_{\rm tip}\,=\,15\,$nm. Considering the minimum distance between the tip apex and the sample surface to be 5\,nm (this value is determined by the setpoint that controls the tip-sample distance in TERS experiments), the shortest distance $z_{\rm min}$ between the graphene surface and the tip dipole becomes $z_{\rm min}=20\,$nm. The curves coincide for $L_{\rm c}=0$, which corresponds to the case where there is no correlation between neighboring graphene lattice points. In this case, the Raman signal is added up incoherently, that is, the intensities of neighboring lattice points are summed up. For a finite correlation length $L_{\rm c}$  we observe that the distance curves for the G and G$^{\prime}$ bands become different, with the G band showing a markedly weaker distance dependence. This deviation arises from the coherent interaction between neighboring lattice points. As $L_{\rm c}$ increases, the distance dependence of the G$^{\prime}$ band transits from $z^{-10}$ to $z^{-8}$. The predictions shown in Fig.~\ref{results_TST} indicate that the phonon correlation length in graphene can be experimentally determined by measuring the G and G$^{\prime}$ approach curves.\\

\begin{figure}
\includegraphics[scale=1]{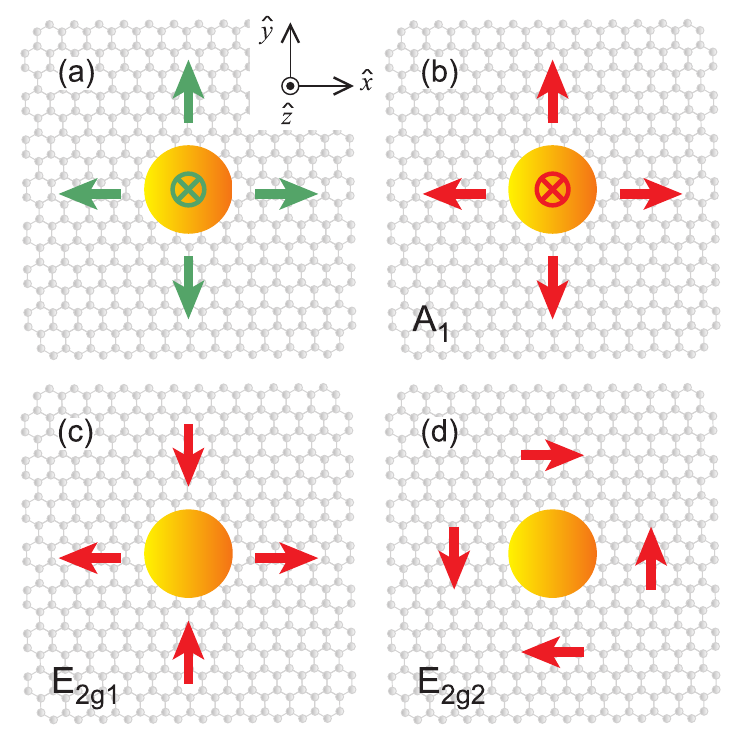}
\begin{narrowtext}
\caption[] {Symmetries of near-field Raman scattering in the {\bf TST} configuration. The yellow spots represent the top view of the tip (axis along the $z$ direction) with the graphene lattice underneath (lying in the $xy$ plane). In (a), the $\otimes$ symbol represents the tip-dipole induced by the incident field. The green arrows represent the $x$ and $y$ in-plane components of the electric field generated by the tip-dipole. The respective induced Raman dipoles in the graphene plane are represented by the red arrows in (b-d) for the modes with A$_{\rm 1}$, E$_{\rm 2g1}$, and E$_{\rm 2g2}$ symmetries, respectively. By considering the fully coherent case ($L_{\rm c}\rightarrow\infty$), the scattered field generated by the A$_{\rm 1}$ Raman dipoles add constructively at the tip apex, generating a strong induced dipole at the tip [represented by the $\otimes$ symbol in (b)]. On the other hand, the field generated by the Raman dipoles interfere destructively at the tip apex for the E$_{\rm 2g1}$ and E$_{\rm 2g2}$ symmetries [(c) and (d), respectively].\label{interference}}
\end{narrowtext}
\end{figure}

The interference effects that generate different enhancements for the G and G$^{\prime}$ bands at finite values of $L_{\rm c}$ are illustrated in Figure~\ref{interference}. The yellow spot represents the top view of the tip (axis along the $z$ direction) with the graphene lattice underneath (lying in the $xy$ plane). In (a), the $\otimes$ symbol represents the tip-dipole induced by the incident field ${\bf E}_{0}(z,\omega)$. The green arrows represent the $x$ and $y$ in-plane components of the electric field ${\bf E}({\bf r};\omega)$ generated by the tip-dipole. The respective induced Raman dipoles ${\bf p}^{\gamma}({\bf r};\omega_{\rm s})$ [Eq.~(\ref{ramandipole})] in the graphene plane are represented by the red arrows in (b-d) for the modes with A$_{\rm 1}$, E$_{\rm 2g1}$, and E$_{\rm 2g2}$ symmetries, respectively. The directions of the induced Raman dipoles are determined by the Raman polarizability tensors $\tensor{\alpha}^{\,\gamma}({\bf r};\omega_{\rm s},\omega)$ in (\ref{alpharaman}). By considering the fully coherent case ($L_{\rm c}\rightarrow\infty$), the scattered field generated by the A$_{\rm 1}$ Raman dipoles add constructively at the tip apex, generating a strong induced dipole at the tip [represented by the $\otimes$ symbol in (b)]. In the other hand, the field generated by the Raman dipoles interfere destructively at the tip apex for the E$_{\rm 2g1}$ and E$_{\rm 2g2}$ symmetries [(c) and (d), respectively]. For finite $L_{\rm c}$, the G band (E$_{\rm 2g}$ symmetry) {\bf TST} signal will not be exactly null, but it will be clearly weaker than in the G$^{\prime}$ case (A$_{\rm 1}$ symmetry). Notice that these two distinct situations are solely associated with the symmetry of the Raman modes.\\

\textbf{ST/2-D:}\\

\begin{figure}
\includegraphics[scale=0.8]{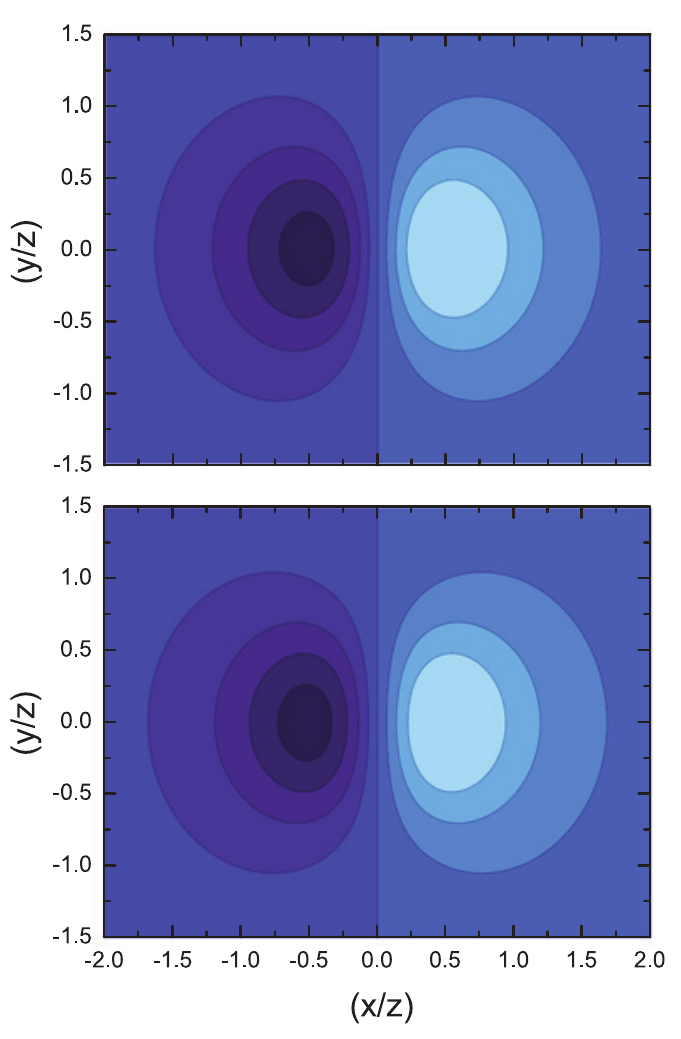}
\vspace{-1em}
\begin{narrowtext}
\caption[] {Comparison of original function (top) and approximation (bottom) according to (\ref{fttrf10}), for $j=x$. The following fitting constants were used:  $a^{\,\prime}_0=0.78$, $b^{\,\prime}_0=2.4$, $c^{\,\prime}_0=0.18$, and $d^{\,\prime}_0=0.56$. \label{Fit_ST}}
\end{narrowtext}
\end{figure}

Now we evaluate the {\bf ST} component [third term in (\ref{excfield4c})] for the 2-D modes (G and G$^{\prime}$). From (\ref{ramanintensity4mn}), the Raman intensity in the {\bf ST} configuration can be calculated as\\
\begin{eqnarray}
S^{\rm ST}({\bf r}_{0}) &=& 4\pi^2 \frac{\omega_{\rm s}^{4}}{\varepsilon_0^2 c^{4}}\,\sum_{l,m,n}\,\sum_{i,j}\;
\tilde\alpha^{\gamma\,\ast}_{mi}\,
\tilde \alpha^{\gamma}_{nj}\,\times \label{ramanintensityTS}\\
&&
 \int\!\!\!\int^{+\infty}_{-\infty} \!\!\!\!\! dk_x \,dk_y\;  \hat{F}^{\ast}_{lmi}(k_x,k_y)\,  \hat{F}_{lnj}(k_x,k_y)\;
\;{\rm e}^{-({k_x}^{\!2}+{k_y}^{\!2}) L_{\rm c}^2/4} \, , \nonumber
\end{eqnarray}\\
where, according to (\ref{excfield4c}), the $\hat{F}^{\ast}_{lnj}(k_x,k_y)$ function (\ref{ftransf09}) in the {\bf ST} configuration assumes the form\\
\begin{eqnarray}
\hat{F}_{lnj}(k_x,k_y) &=& \frac{1}{4\pi^2} \,\frac{\omega^{2}}{\varepsilon_0 c^{2}}\,
\alpha_{\parallel}(\omega)\;\!\,{\rm G}_{ln}^{\rm o}({\bf r}_{0},z;\omega_{\rm s})\,E_{0\,z}(z,\omega) \label{ftransfTS}
 \\[0ex]
 &&\hspace{5em} \times
\int\!\!\!\int^{+\infty}_{-\infty} \!\!\!\!\! dx \,dy
\,{\rm G}_{jz}^{\rm o}(x,y,z;\omega)\,
 {\rm e}^{-i(k_x x + k_y y)} \;
\nonumber \\
&& = \frac{3}{16\pi^{3}\,\varepsilon_0}\alpha_{\parallel}(\omega)\,{\rm G}_{ln}^{\rm o}({\bf r}_{0},z;\omega_{\rm s})\,E_{0\,z}(z,\omega) \nonumber
 \\[0ex]
 &&\hspace{5em} \times
\int\!\!\!\int^{+\infty}_{-\infty} \!\!\!\!\! dx \,dy
\,\frac{j z}{(x^2+y^2+z^2)^{(5/2)}}\,
 {\rm e}^{-i(k_x x + k_y y)}  \; .\nonumber
\end{eqnarray}\\
As before, the integrand in (\ref{ftransfTS}) can be approximated as\\
\begin{eqnarray}
\frac{j z}{(x^2+y^2+z^2)^{(5/2)}}\;\approx \;\frac{j}{z^4}\,\left[ a^{\prime}_0\,{\rm e}^{-b^{\prime}_0 \;\!(x^2 + y^2)/z^2}\,+\,c^{\prime}_0\,{\rm e}^{-d^{\prime}_0 \;\!(x^2 + y^2)/z^2}\right] \, ,
\label{fttrf10}
\end{eqnarray}\\
with $a^{\,\prime}_0$, $b^{\,\prime}_0$, $c^{\,\prime}_0$, and $d^{\,\prime}_0$ being fitting constants. Figure~\ref{Fit_ST} shows a comparison of the original function and the approximation.\\

\begin{figure}
\includegraphics[scale=0.90]{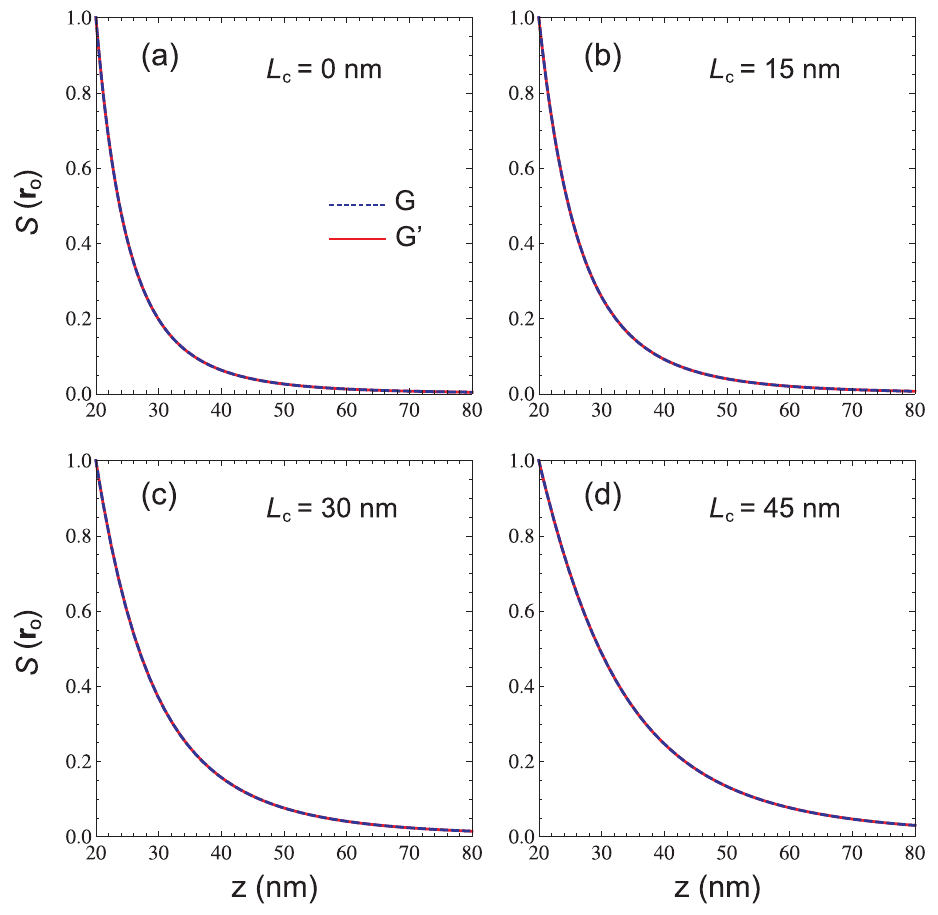}
\begin{narrowtext}
\caption[] {Distance dependence of the {\bf ST} signal for the G and G$^{\prime}$ bands. Panels (a-d) account for different values of the correlation length $L_{\rm c}$: 0, 15, 30, and 45\,nm, respectively, as indicated in the graphics. In all cases we used $\tilde{f}_{\rm e}=3$ and $r_{\rm tip}=15\,$nm. The signal $S({\bf r}_{0})$ is normalized to $1$ at $z_{\rm min}=20\,$nm. Compared to the {\bf TST} signal shown in Fig.~\ref{results_TST}, the {\bf ST} signal presents a weaker decay and therefore contributes to the measured signal only for larger tip-sample distances.\label{results_ST}}
\end{narrowtext}
\end{figure}

The Fourier transform (\ref{ftransfTS}) can now be evaluated as\\
\begin{equation}
\hat{F}_{lnj}(k_x,k_y)\,=\,\frac{(-i)\,3}{32\,\pi^2 \,\varepsilon_0}\,\alpha_{\parallel}(\omega)\,{\rm G}_{ln}^{\rm o}({\bf r}_{0},z;\omega_{\rm s})\,E_{0\,z}(z,\omega)\, \hat{h}_{j}(k_x,k_y;z)\;,\label{ftransf20}
\end{equation}\\
with\\
\begin{equation}
\hat{h}_{j}(k_x,k_y;z)=k_{j}\bigg[\frac{a^{\,\prime}_0}{b_0^{\,\prime\,2}\,} {\rm e}^{-(k_x^2 + k_y^2) z^2/4 b^{\,\prime}_{0}}+\,\frac{c^{\,\prime}_{0}}{d_0^{\,\prime\,2}} {\rm e}^{- (k_x^2 + k_y^2) z^2/4 d^{\,\prime}_{0}}\bigg]\,.\label{ftranfts}
\end{equation}\\
To calculate the Raman signal (\ref{ramanintensityTS}) for the \textbf{ST} interaction term we define the expressions \\
\begin{eqnarray}
\ell_{ij}(z,L_{\rm c})\;=\;  \int\!\!\!\int^{+\infty}_{-\infty} \!\!\!\!\! dk_x \,dk_y\;  \hat{h}^{\ast}_{i}(k_x,k_y;z)\,  \hat{h}_{j}(k_x,k_y;z)\;
\;{\rm e}^{-({k_x}^{\!2}+{k_y}^{\!2}) L_{\rm c}^2/4} \,,\label{ellij}
 \end{eqnarray}\\
which can be solved analytically. They have the properties\\
\begin{eqnarray}
\ell_{ij}(z, L_{\rm c}\rightarrow\infty) = 1 / L_{\rm c}^4 \,,\qquad \ell_{ij}(z, L_{\rm c}\rightarrow 0) = 1/z^4 \; .
\end{eqnarray}\\
Inserting (\ref{ftransf20}-\ref{ellij}) into (\ref{ramanintensityTS}) yields\\
\begin{eqnarray}
S^{\rm ST}({\bf r}_{0}) &=& \frac{9}{256 \pi^2}\,\frac{\omega_{\rm s}^{4}}{\varepsilon_0^4 c^{4}}\,
\sum_{l,m,n}\,\sum_{i}\,\left[{\rm G}_{lm}^{\rm o\,\ast}({\bf r}_{0},z;\omega_{\rm s}){\rm G}_{ln}^{\rm o}({\bf r}_{0},z;\omega_{\rm s})\left|\alpha_{\parallel}(\omega)\,E_{0\,z}(z,\omega)\right|^{2}\right] \nonumber\\
&&  \hspace{10em}\;\; \times \left[ \tilde\alpha^{\gamma\,\ast}_{mi}\,
\tilde \alpha^{\gamma}_{ni} \, \ell_{ii}(z,L_{\rm c})\, \right], \label{reslt01TS}
\end{eqnarray}\\
where we have used the properties $\ell_{xx}(z,L_{\rm c})=\ell_{yy}(z,L_{\rm c})\neq\,0$, and $\ell_{xy}(z,L_{\rm c})=\ell_{yx}(z,L_{\rm c})=0$. For the modes with E$_{\rm 2g1}$ and A$_{1}$ symmetries, the {\bf ST} component of the signal is given as\\
\begin{eqnarray}
S^{\rm ST}({\bf r}_{0})({\rm E_{2g1},A_{1}}) &=& \frac{9}{64}\,\frac{\omega_{\rm s}^{4}}{\varepsilon_0^4 c^{4}}\,
\sum_{l}\left|{\rm G}_{lx}^{\rm o}({\bf r}_{0},z;\omega_{\rm s})\alpha_{\parallel}(\omega)\,E_{0\,z}(z,\omega)\right|^{2}\nonumber\\
&&  \hspace{4em}\;\; \times \left(\tilde\alpha^{\gamma\,\ast}_{xx}\tilde\alpha^{\gamma}_{xx}\,+\,\tilde\alpha^{\gamma\,\ast}_{yy}\tilde\alpha^{\gamma}_{yy}\right) \, \ell_{xx}(z,L_{\rm c}), \label{reslt02TS}
\end{eqnarray}\\
where, based on the radial symmetry of the system, we have considered ${\rm G}_{lx}^{\rm o}={\rm G}_{ly}^{\rm o}$. For the E$_{\rm 2g2}$ symmetry, the {\bf ST} component assumes the form\\
\begin{eqnarray}
S^{\rm ST}({\bf r}_{0})({\rm E_{2g2}}) &=& \frac{9}{64}\,\frac{\omega_{\rm s}^{4}}{\varepsilon_0^4 c^{4}}\,
\sum_{l}\left|{\rm G}_{lx}^{\rm o}({\bf r}_{0},z;\omega_{\rm s})\alpha_{\parallel}(\omega)\,E_{0\,z}(z,\omega)\right|^{2}\nonumber\\
&&  \hspace{4em}\;\; \times \left(\tilde\alpha^{\gamma\,\ast}_{xy}\tilde\alpha^{\gamma}_{xy}\,+\,\tilde\alpha^{\gamma\,\ast}_{yx}\tilde\alpha^{\gamma}_{yx}\right) \, \ell_{xx}(z,L_{\rm c}). \label{reslt03TS}
\end{eqnarray}\\
Expressing $\alpha_{\parallel}$ in terms of the complex field-enhancement factor $f_{\rm e}(\omega)$  and the tip radius $r_{\rm tip}$  [Eq.~(\ref{tip03})], and inserting the Raman tensor components [Eq.~(\ref{alpharaman})] into (\ref{reslt02TS}) and (\ref{reslt03TS}), we obtain the following expressions for the G and G$^{\prime}$ Raman signals\\
\begin{eqnarray}
S_{\rm G}^{\rm ST}({\bf r}_{0}) &=& \frac{9}{16}\,\frac{\omega_{\rm s}^{4}}{\varepsilon_0^2 c^{4}}\,r_{\rm tip}^{6}\,\tilde{f}_{\rm e}(\omega)^{2}\,
\sum_{l}\left|{\rm G}_{lx}^{\rm o}({\bf r}_{0},z;\omega_{\rm s})\,E_{0\,z}(z,\omega)\,\tilde\alpha^{\rm G}(\omega_{\rm s},\omega)\right|^{2}\,\ell_{xx}(z,L_{\rm c}), \label{resltTSG}
\end{eqnarray}\\
 \begin{eqnarray}
S_{\rm G'}^{\rm ST}({\bf r}_{0}) &=& \frac{9}{32}\,\frac{\omega_{\rm s}^{4}}{\varepsilon_0^2 c^{4}}\,r_{\rm tip}^{6}\,\tilde{f}_{\rm e}(\omega)^{2}\,
\sum_{l}\left|{\rm G}_{lx}^{\rm o}({\bf r}_{0},z;\omega_{\rm s})\,E_{0\,z}(z,\omega)\,\tilde\alpha^{\rm G'}(\omega_{\rm s},\omega)\right|^{2}\,\ell_{xx}(z,L_{\rm c}). \label{resltTS2D}
\end{eqnarray}\\
Equations (\ref{resltTSG}) and (\ref{resltTS2D}) show that, unlike the {\bf TST} case, the only differences between the G and G$^{\prime}$ {\bf ST} signals are their numerical pre-factors and Raman efficiencies, expressed in terms of $\tilde\alpha^{\rm G}$ and $\tilde\alpha^{\rm G'}$. Figure \ref{results_ST} shows the plot of the distance dependence of the G and G$^{\prime}$ {\bf ST} signals [according to Eqs. (\ref{resltTSG}) and (\ref{resltTS2D}), respectively] for different correlation lengths $L_{\rm c}$, assuming $r_{\rm tip}\,=\,15\,$nm, and $\tilde{f}_{\rm e}=3$. The signal is normalized to 1 at $z_{\rm min}=20$\,nm. As expected, the G and G$^{\prime}$ curves coincide for all values of $L_{\rm c}$. For $L_{\rm c}=0$, they show a dependence to $z^{-4}$. For a finite correlation length $L_{\rm c}$, we observe a slight drop in the $z$ dependence. For $L_{\rm c}\rightarrow\infty$, both of them show no enhancement.\\

\textbf{TST/1-D:}\\

Next, we evaluate the {\bf TST} component of the near-field signal for one-dimensional (1-D) systems, more specifically for the D band at graphene edges. The D mode is totally symmetric (A$_{\rm 1}$ symmetry), and we consider the edge along the $x$-direction, with coordinate $y=0$. In this case, the position vector at the sample is reduced to ${\bf r}=(x,0,0)$. An important factor to be taken into account is the strong depolarization effect in the optical absorption of 1-D systems, for which the absorption is maximum for light polarized along the longitudinal direction of the object, and null for light polarized along its transverse direction~\cite{casiraghi09b,cancado04}. To account for depolarization, we introduce a depolarization tensor $\tensor{d}$ with which the local excitation field [Eq.~(\ref{excfield})] becomes\\
\begin{equation}
{\bf E}_{\rm tot}(x,\omega) = \tensor{d}\left[
{\bf E}_0(x,\omega)\,+\,\frac{\omega^{2}}{\varepsilon_0 c^{2}} \:
\tensor{\rm G}^{\rm o}(x,z;\omega)\, \tensor{\alpha}_{\rm tip}(\omega)\,{\bf E}_0(z,\omega)\right]\,,
\label{totalfield}
\end{equation}\\
with\\
\begin{eqnarray}
\tensor{d}(\omega)\left[
\begin{array}{rrr}
1 & 0 & 0 \\
0 & 0 & 0 \\
0 & 0 & 0
\end{array}
\right]\,.
\label{depolarization}
\end{eqnarray}\\
The Raman induced dipole [Eq.~(\ref{ramandipole})] is now evaluated as\\
\begin{equation}
{\bf p}^{\gamma}(x,\omega_{\rm  s}) \,=\,
\tensor{\alpha}^{\,\gamma}(x,\omega_{\rm  s};\omega) \,
{\bf E}_{\rm tot}(x,\omega) \;.
\label{dipole}
\end{equation}\\
The same depolarization effect accounts for the scattered field and, in this case, Eq.~(\ref{scattereddipole}) becomes\\
\begin{equation}
{\bf E}({\bf r}_{0},\omega_{s}) \,=\, \tensor{d}\,\left[\frac{\omega_{\rm s}^{2}}{\varepsilon_0 c^{2}}
\int^{+\infty}_{-\infty} \!\!\!\!\! dx\;\tensor{\rm G}({\bf r}_{0},x;\omega_{\rm s}) \,
{\bf p}^{\gamma}(x;\omega_{\rm s})\,\right] \,.
\label{efield1D}
\end{equation}\\

\begin{figure}
\includegraphics[scale=0.9]{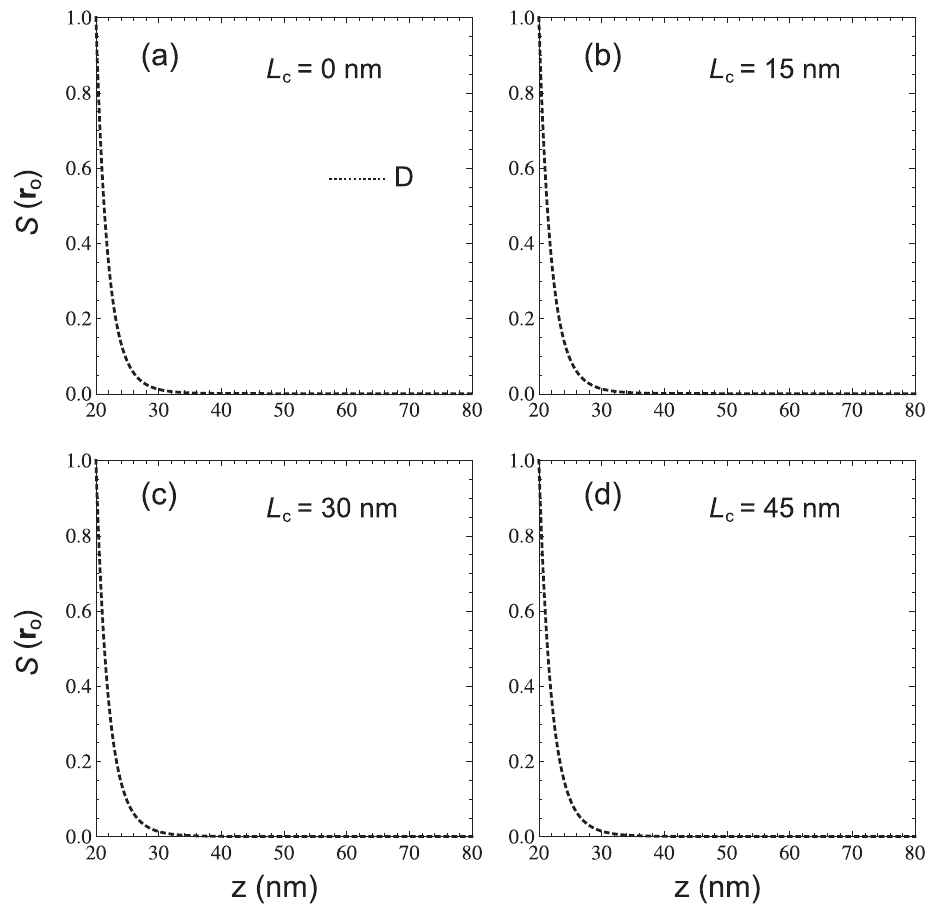}
\begin{narrowtext}
\caption[] {Dependence of the {\bf TST} of the D band on tip-sample separation $z$. Panels (a-d) account for different values of $L_{\rm c}$: 0, 15, 30, and 45\,nm, respectively, as indicated in the graphics. In all cases we used $\tilde{f}_{\rm e}=3$ and $r_{\rm tip}=15\,$nm. The signal $S({\bf r}_{0})$ is normalized to $1$ at $z_{\rm min}=20\,$nm.\label{results_TST_1D}}
\end{narrowtext}
\end{figure}

For totally symmetric modes in 1-D systems, the scattered signal in the {\bf TST} configuration [Eq.~(\ref{ramanintensity4mn}) for the 2-D case] is given by\\
\begin{eqnarray}
S({\bf r}_{0}) &=& 2\pi \frac{\omega_{\rm s}^{4}}{\varepsilon_0^2 c^{4}}\,
 \sum_{l}\;\tilde\alpha^{\gamma\,\ast}_{xx}\,
\tilde \alpha^\gamma_{xx}
 \int^{+\infty}_{-\infty} \!\!\!\!\! dk_x\;  \hat{F}^{\ast}_{lx}(k_x)\,  \hat{F}_{lx}(k_x)\;
\;{\rm e}^{-(k_{x}^2 L_{\rm c}^2)/4} \,.
\label{signal1D}
 \end{eqnarray}\\
The Fourier components $\hat{F}^{\ast}_{lx}(k_x)$ in (\ref{signal1D}) can be evaluated as\\
\begin{eqnarray}
\hat{F}_{lx}(k_x) &=& \frac{1}{2\pi} \,\frac{\omega^{2}\omega_{\rm s}^2}{\varepsilon_0^2 c^{4}}\,
\alpha_{\parallel}(\omega)\;\!\alpha_{\parallel}(\omega_{\rm s})\,{\rm G}_{lz}^{\rm o}({\bf r}_{0},z;\omega_{\rm s})\,E_{0\,z}(z,\omega) \label{ftransf1D}
 \\[0ex]
 &&\hspace{5em} \times
\int^{+\infty}_{-\infty} \!\!\!\!\! dx
\,{\rm G}_{zx}^{\rm o}(x,y,z;\omega_{\rm s}) \,{\rm G}_{xz}^{\rm o}(x,y,z;\omega)\,
 {\rm e}^{-ik_x x} \; .
\nonumber
\end{eqnarray}\\
Using the same approximations as for the {\bf TST} scattering in 2-D systems [see Eq.~(\ref{fttrf09})], we obtain\\
\begin{equation}
\hat{F}_{lx}(k_x)\,=\,\frac{9}{128\,\pi^{5/2} \,\varepsilon_0^2\, z^{5}}\,
\alpha_{\parallel}(\omega)\;\!\alpha_{\parallel}(\omega_{\rm s})\,{\rm G}_{lz}^{\rm o}({\bf r}_{0},z;\omega_{\rm s})\,E_{0\,z}(z,\omega)\, \hat{h}_{x}(k_x;z)\;,\label{ftransf21D}
\end{equation}\\
where\\
\begin{eqnarray}
&\hat{h}_{x}(k_x;z)=&\bigg[\frac{a_0 (2b_0 - k_x^2 z^2)}{b_0^{\,5/2}} {\rm e}^{- (k_x^2 z^2)\,/\, 4 b_0} \,+\,\frac{c_0 (2d_0 - k_x^2 z^2)}{d_0^{\,5/2}} {\rm e}^{- (k_x^2 z^2)\,/\, 4 d_0}\bigg]\, ,
\end{eqnarray}\\
with $a_0=0.74$, $b_0=4.0$, $c_0=0.08$, and $d_0=1.5$ (same values as obtained for the {\bf TST} scattering in 2-D systems). To calculate the Raman signal (\ref{signal1D}) we define the expression \\
\begin{eqnarray}
\ell_{xx}(z,L_{\rm c})\;=\;\int^{+\infty}_{-\infty} \!\!\!\!\! dk_x \;  \hat{h}^{\ast}_{x}(k_x;z)\,\hat{h}_{x}(k_x;z)\;
\;{\rm e}^{-(k_{x}^{2}\,L_{\rm c}^2)/4} \,, \label{fminj1D}
 \end{eqnarray}\\
which can be solved analytically. It has the properties\\
\begin{eqnarray}
\ell_{xx}(z, L_{\rm c}\rightarrow\infty) = 1 / L_{\rm c}\;,\qquad \ell_{xx}(z, L_{\rm c}\rightarrow 0) = 1/z \; .
\end{eqnarray}\\
Inserting (\ref{fminj1D}) into (\ref{signal1D}) yields\\
\begin{equation}
S^{\rm TST}({\bf r}_{0}) =\frac{81}{512}\,\frac{\omega_{\rm s}^{4}}{\varepsilon_0^2 c^{4}}\frac{r_{\rm tip}^{12}\,\tilde{f}_{\rm e}(\omega)^{4}}{z^{10}}\,
\sum_{l}\;\left|{\rm G}_{lz}^{\rm o}({\bf r}_{0},z;\omega_{\rm s})\,E_{0\,z}(z,\omega)\: \tilde\alpha^{\rm D}(\omega_{\rm s};\omega)
 \right|^2 \,\ell_{xx}(z,L_{\rm c})\,,  \quad \label{resltTST1D}
 \end{equation}\\
where we have used $\tilde \alpha^\gamma_{xx}=1$, according to (\ref{alpharaman}). We find that for a 1-D system with zero correlation length ($L_{\rm c}\rightarrow 0$) the {\bf TST} signal from a totally symmetric mode decays as $z^{-11}$, whereas for infinite correlation length ($L_{\rm c}\rightarrow \infty$) it decays as $z^{-10}$, consistent with the theory described in Ref.~\cite{cancado2009}. Figure~\ref{results_TST_1D} shows the plot of the distance dependence of the {\bf TST} signal for the D band [according to Eq.~(\ref{resltTST1D})] for different correlation lengths $L_{\rm c}$, assuming $r_{\rm tip}\,=\,15\,$nm, and $\tilde{f}_{\rm e}=3$. The signal is normalized to 1 at $z_{\rm min}=20$\,nm. The smaller the correlation length is, the steeper is the distance dependence.\\

\textbf{ST/1-D:}\\

\begin{figure}
\includegraphics[scale=0.9]{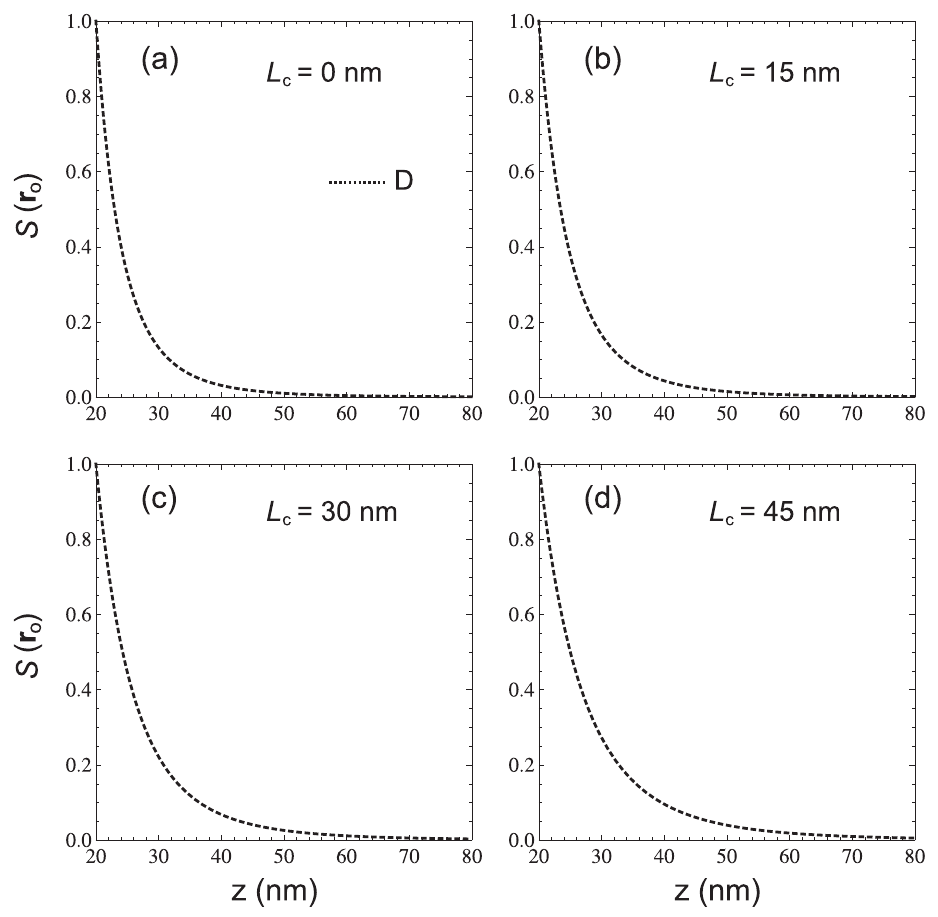}
\begin{narrowtext}
\caption[] {Distance dependence of the {\bf ST} signal for the D band. Panels (a-d) account for different values of $L_{\rm c}$: 0, 15, 30, and 45\,nm, respectively, as indicated in the graphics. In all cases we set $\tilde{f}_{\rm e}=3$ and $r_{\rm tip}=15\,$nm. The signal $S({\bf r}_{0})$ is normalized to $1$ at $z_{\rm min}=20\,$nm. \label{results_ST_1D}}
\end{narrowtext}
\end{figure}

Starting from Eq.~(\ref{ramanintensity4mn}), the {\bf ST} component of the scattered signal for the 1-D case can be calculated as\\
\begin{equation}
S^{\rm ST}({\bf r}_{0}) = 2\pi \frac{\omega_{\rm s}^{4}}{\varepsilon_0^2 c^{4}}\,
 \sum_{l}\tilde\alpha^{\gamma\,\ast}_{xx}\,
\tilde \alpha^\gamma_{xx}\,\int^{+\infty}_{-\infty} \!\!\!\!\! dk_x;  \hat{F}^{\ast}_{lx}(k_x)\,  \hat{F}_{lx}(k_x)\;
\;{\rm e}^{-(k_{x}^{2}L_{\rm c}^{2})/4} \, , \label{ramanintensityTS1D}
 \end{equation}\\
where, according to (\ref{excfield4c}), the Fourier $\hat{F}_{lx}(k_x)$ component has the form\\
\begin{equation}
\hat{F}_{lx}(k_x)= \frac{1}{2\pi} \,\frac{\omega^{2}}{\varepsilon_0 c^{2}}\,
\alpha_{\parallel}(\omega)\;\!\,{\rm G}_{lx}^{\rm o}({\bf r}_{0},z;\omega_{\rm s})\,E_{0\,z}(z,\omega)\,\int^{+\infty}_{-\infty} \!\!\!\!\! dx \,{\rm G}_{xz}^{\rm o}(x,z;\omega)\,
 {\rm e}^{-i\,k_x x} \;. \label{ftransfTS1D}
\end{equation}\\
Considering the same approximations as for the 2-D case [Eq.~(\ref{fttrf10})], the Fourier component (\ref{ftransfTS1D}) can be evaluated as\\
\begin{equation}
\hat{F}_{lx}(k_x)\,=\,\frac{(-i)\,3}{16\,\pi^{3/2} \,\varepsilon_0\, z}\,\alpha_{\perp}(\omega)\,{\rm G}_{lx}^{\rm o}({\bf r}_{0},z;\omega_{\rm s})\,E_{0\,z}(z,\omega)\, \hat{h}_{x}(k_x)\;,\label{ftransf201D}
\end{equation}\\
with\\
\begin{equation}
\hat{h}_{x}(k_x)=k_{x}\bigg[\frac{a^{\,\prime}_0}{{b_0}^{\prime\,3/2}\,}{\rm e}^{-(k_x z)^2/4b^{\,\prime}_{0}}+\,\frac{c^{\,\prime}_{0}}{d_0^{\,\prime\,3/2}} {\rm e}^{- (k_x z)^2/4d^{\,\prime}_{0}}\bigg]\,,\label{ftranftsTS1D}
\end{equation}\\
where the fitting parameters $a^{\,\prime}_0$, $b^{\,\prime}_0$, $c^{\,\prime}_0$, and $d^{\,\prime}_0$ are the same as obtained for the 2-D case ($a^{\,\prime}_0=0.78$, $b^{\,\prime}_0=2.4$, $c^{\,\prime}_0=0.18$, and $d^{\,\prime}_0=0.56$). We introduce the function\\
\begin{eqnarray}
\ell_{xx}(z,L_{\rm c})\;=\;\int^{+\infty}_{-\infty} \!\!\!\!\! dk_x\;  \hat{h}^{\ast}_{x}(k_x)\,  \hat{h}_{x}(k_x)\;
\;{\rm e}^{-(k_{x}^{\!2}L_{\rm c}^{2})/4} \,,\label{ellij1D}
 \end{eqnarray}\\
which can be calculated analytically and has the properties\\
\begin{eqnarray}
\ell_{xx}(z, L_{\rm c}\rightarrow\infty) = 1 / L_{\rm c}^3\,,\qquad \ell_{xx}(z, L_{\rm c}\rightarrow 0) = 1/z^3 \; .
\label{lcST}
\end{eqnarray}\\
Inserting (\ref{ftransf201D}-\ref{ellij1D}) into (\ref{ramanintensityTS1D}) yields\\
\begin{eqnarray}
S^{\rm ST}({\bf r}_{0}) &=& \frac{9}{32}\,\frac{\omega_{\rm s}^{4}}{\varepsilon_0^2 c^{4}\,z^2}\,r_{\rm tip}^{6}\,\tilde{f}_{\rm e}(\omega)^{2}\,
\sum_{l}\left|{\rm G}_{lx}^{\rm o}({\bf r}_{0},z;\omega_{\rm s})\,E_{0\,z}(z,\omega)\,\tilde\alpha_{0}^{\rm G}(\omega_{\rm s},\omega)\right|^{2}\,\ell_{xx}(z,L_{\rm c}), \label{resltTS1D}
\end{eqnarray}\\
where we have used $\tilde \alpha^\gamma_{xx}=1$. According to (\ref{lcST}) and (\ref{resltTS1D}), for $L_{\rm c}=0$, the D band signal in the {\bf ST} configuration is proportional to $z^{-5}$, and for $L_{\rm c}\rightarrow\infty$ the signal is proportional to $z^{-2}$ (see Figure~\ref{results_ST_1D}). Similar to the 2-D case for the Raman G and G$^{\prime}$ bands, we find that the {\bf ST} term has a weaker distance dependence than the {\bf TST} term and therefore it contributes to the measured signal only for large tip-sample distances.\\

Finally, we summarize our findings and discuss the main results. We have presented the theory of near-field Raman scattering accounting for spatial source correlations and associated coherence properties of the scattered signal. Our calculations were performed for a TERS configuration, where a metal tip acting as an optical antenna is positioned near the sample. We considered the {\bf TST} and {\bf ST} components of the scattered field, and the calculations were performed for one-dimensional (1-D) and two-dimensional (2-D) samples. The theory was applied specifically to graphene, namely the D, G, and G$^{\prime}$ Raman bands. While the D and G$^{\prime}$ bands are associated with totally symmetric (A$_{1}$) phonons, the G band originates from a double degenerate mode with E$_{\rm 2g}$ symmetry. On the other hand, while the G and G$^{\prime}$ bands are allowed over the whole graphene area, the D band is strongly localized at the edges, which define a one-dimensional sample.\\

\begin{figure}
\includegraphics[scale=0.9]{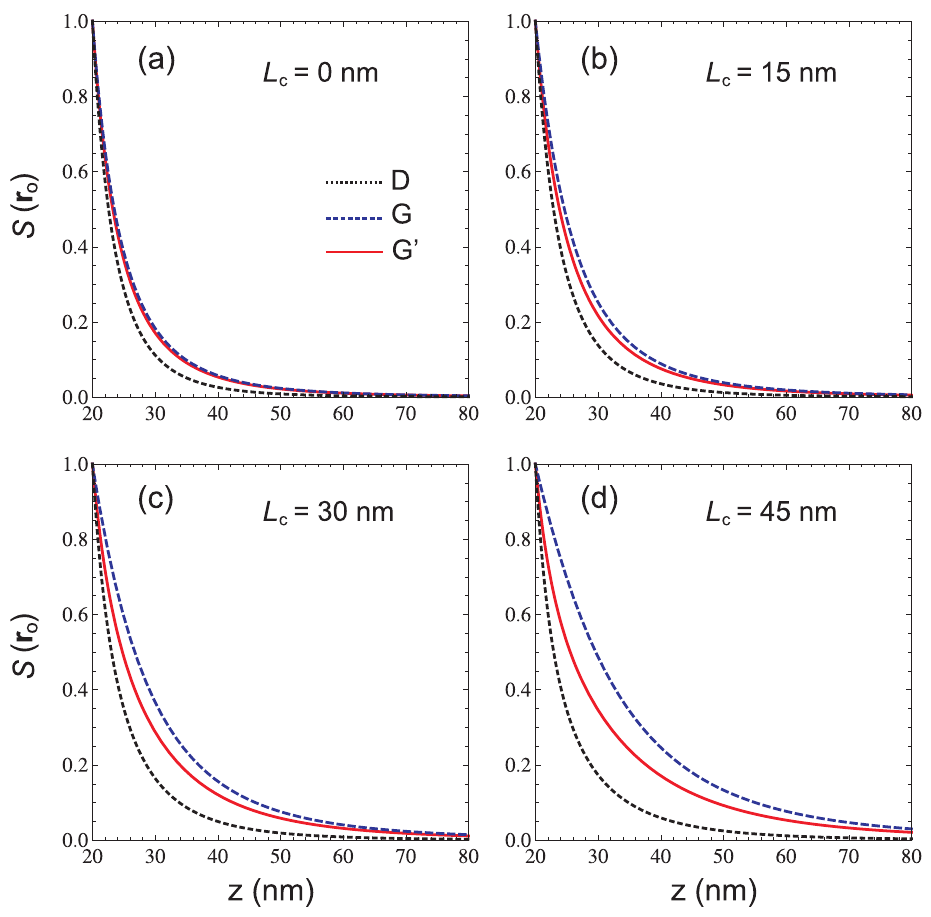}
\begin{narrowtext}
\caption[] { Distance dependence of the {\bf TST}+{\bf ST} signals for the D, G and G$^{\prime}$ bands. Panels (a-d) account for different values of $L_{\rm c}$: 0, 15, 30, and 45\,nm, respectively, as indicated in the graphics. In all cases we used $\tilde{f}_{\rm e}=3$ and $r_{\rm tip}=15\,$nm. The signal $S({\bf r}_{\rm 0})$ is normalized to $1$ at $z_{\rm min}=20\,$nm. \label{results_TST+ST}}
\end{narrowtext}
\end{figure}

For samples with finite correlation length $L_{\rm c}$, the {\bf TST} term gives rise to a characteristic difference between the G and G$^{\prime}$ Raman signals. This difference is associated with near-field interferences which, in the case of the Raman G band, are destructive, and in the case of the G$^{\prime}$  band, turn out to be constructive (see Fig.~\ref{interference} and associated discussion). The near-field interferences give rise to a weaker tip-sample distance dependence for the G band than for the G$^{\prime}$ band, and make it possible to extract the correlation length $L_{\rm c}$ from measured data. Moreover, when the D band signal originates from the edges (1-D geometry), a further modification of the distance dependence is observed. All these effects are summarized in Figure~\ref{results_TST+ST}, which shows the plot of the {\bf TST}+{\bf ST} signal versus the tip-sample separation ($z$) for different values of $L_{\rm c}$  (0, 15, 30, and 45\,nm). In all cases we used $\tilde{f}_{\rm e}=3$ and $r_{\rm tip}=15$\,nm. The G and 2D curves coincide for $L_{\rm c}=0$, as expected. The D band signal presents a different trend, showing a steeper tip-sample distance dependence (stronger enhancement). For finite values of $L_{\rm c}$, the distance dependence drops for all bands as $L_{\rm c}$ increases. Simultaneously, G and G$^{\prime}$ distance curves become different, with the G band showing a markedly weaker enhancement. Therefore, the experimental observation of different tip-sample distance dependencies for the G and G$^{\prime}$ bands provides strong evidence for near-field interference effects associated with finite correlation lengths $L_{\rm c}$. \\

The curves shown in Fig.~\ref{results_TST+ST} were reproduced in experimental measurements in Ref.~\cite{beams2014}, where we have measured the near-field Raman signal of the D, G and G$^{\prime}$ bands on graphene samples. In these experiments, the correlation length of the G and D optical phonons were obtained by fitting the near-field experimental data with the theory presented here. For both cases (D and G phonons) we found $L_{\rm c}\approx30$\,nm. Note that it is possible to extract $L_{\rm c}$ from the width of the Raman lines in nano-structured systems~\cite{ley1981}, but the obtained value is an average over the laser-irradiated sample area. On the other hand, the near-field procedure developed here allows for the measurement of $L_{\rm c}$ in single crystals and with nanoscale spatial resolutions, which makes it applicable to the analysis of transport properties of a wide range of materials. Most importantly, our work demonstrates that it is not a priori legitimate to treat Raman scattering as an incoherent process in which the signal from different sample regions is simply summed up.\\

We are grateful for valuable input from P. Scott Carney and for financial support by the U. S. Department of Energy (grant DE-FG02-05ER46207) and the Swiss National Science Foundation (grant 200021$\_$149433). AJ acknowledges CAPES for financing his stay at ETH. AJ and LGC acknowledge financial support from CNPq and FAPEMIG.\\

\end{document}